\documentclass[10pt,journal,compsoc]{IEEEtran}
  \usepackage{cite}

\ifCLASSINFOpdf
   \usepackage[pdftex]{graphicx}

\else

\fi

\usepackage{grffile}

\usepackage[nottoc]{tocbibind}

\usepackage{multirow}

\usepackage{ucs}
\usepackage{amsmath}
\usepackage{amsfonts}
\usepackage{amssymb}
\usepackage{makeidx}
\usepackage{breqn} 

\usepackage{url}
\usepackage{hyperref}
\usepackage{listings}
\usepackage{color}

\usepackage[table,xcdraw]{xcolor}
\usepackage[normalem]{ulem}
\useunder{\uline}{\ul}{}

\definecolor{dkgreen}{rgb}{0,0.6,0}
\definecolor{gray}{rgb}{0.5,0.5,0.5}
\definecolor{mauve}{rgb}{0.58,0,0.82}

\usepackage{pgfplots}
\usepackage{calc}
\usepackage{ifthen}
\usepackage{tikz}

\usepackage{graphicx}

\usepackage{rotating}

\begin{document}
\newcommand{\slice}[4]{
	\pgfmathparse{0.5*#1+0.5*#2}
	\let\midangle\pgfmathresult
	
	\draw[thick,fill=black!10] (0,0) -- (#1:1) arc (#1:#2:1) -- cycle;
	
	\node[label=\midangle:#4] at (\midangle:1) {};
	
	\pgfmathparse{min((#2-#1-10)/110*(-0.3),0)}
	\let\temp\pgfmathresult
	\pgfmathparse{max(\temp,-0.5) + 0.8}
	\let\innerpos\pgfmathresult
	\node at (\midangle:\innerpos) {#3};
}

\title{Software Engineers vs.\\ Machine Learning Algorithms: \\
	An Empirical Study Assessing Performance and Reuse Tasks }

\author{
	Nathalia~Nascimento,~Carlos~Lucena, 
\IEEEcompsocitemizethanks{\IEEEcompsocthanksitem Laboratory of Software Engineering (LES) at the Department of Informatics, Pontifical Catholic University of Rio de Janeiro, Brazil. CAPES scholarship/Program 194/Process: 88881.134630/2016-01 \protect\\
E-mail: nnascimento,lucena@inf.puc-rio.br
\IEEEcompsocthanksitem See http://www.inf.puc-rio.br/~nnascimento/ }
Paulo~Alencar~
and ~Donald~Cowan
\IEEEcompsocitemizethanks{\IEEEcompsocthanksitem David R. Cheriton School of Computer Science. University of Waterloo, Canada. \protect\\
	E-mail: palencar,dcowan@csg.uwaterloo.ca
	}
}


\IEEEtitleabstractindextext{%
	\begin{abstract}
		
Several papers have recently contained reports on  applying machine learning (ML) to the automation of software engineering (SE) tasks, such as project management, modeling and development. However, there appear to be no approaches comparing how software engineers fare against machine-learning algorithms as applied to specific software development tasks. Such a comparison is essential to gain insight into  which tasks are better performed by humans and which by machine learning and how  cooperative work or human-in-the-loop processes can be implemented more effectively.  In this paper, we present an empirical study that  compares how software engineers and machine-learning algorithms perform and reuse tasks. The empirical study involves the synthesis of the control structure of an autonomous streetlight application. Our approach consists of four steps. First, we solved the problem using machine learning to determine specific performance and reuse tasks. Second, we asked software engineers with different domain knowledge levels to provide a solution to the same tasks. Third, we compared how software engineers fare against machine-learning algorithms when accomplishing the performance and reuse tasks based on criteria such as energy consumption and safety. Finally, we analyzed the results to understand which tasks are better performed by either humans or algorithms so that they can work together more effectively. Such an understanding and the resulting human-in-the-loop approaches, which take into account the strengths and weaknesses of humans and machine-learning algorithms, are fundamental not only to provide a basis for cooperative work in support of software engineering, but also, in other areas.

	\end{abstract}
	
}



\maketitle

\IEEEdisplaynontitleabstractindextext

\IEEEpeerreviewmaketitle

\IEEEraisesectionheading{\section{Introduction}\label{sec:introduction}}

\IEEEPARstart{S}{oftware}  engineering processes can be very complex, costly and time-consuming \cite{brooks1987no}. They typically consist of a collection of related tasks \cite{pressman2005software} such as designing, implementing, maintaining, testing and reusing software applications \cite{zhang2000software}. In addition, as software has become embedded in systems of all kinds, millions of computer programs have to be corrected, adapted, and enhanced \cite{pressman2005software}. As a result, the field of software engineering requires millions of skilled Information Technology (IT) professionals to create millions of lines of code, which must be installed, configured, tuned, and maintained. According to Kephart (2005) \cite{kephart2005research}, in the near future, it will be extremely challenging to manage IT environments, even for the most skilled IT professionals.

Several researchers have proposed the use of artificial intelligence, especially machine-learning (ML) techniques, to automate different software engineering (SE) tasks \cite{mostow1985foreword, barstow1987artificial,partridge1988artificial,cheung1991survey, partridge1998artificial, van1998inferring, zhang2000software, boetticher2001using,padberg2004using,zhang2000applying,zhang2006machine,zhang2005machine,zhang2008machine,khoshgoftaar2003introduction,zhang2009machine,zhang2002machine, kramer2000gaps}. For example, Zhang has extensively studied this theme recently and in \cite{zhang2000software} he stated that:

\begin{quote}
	''The field of software engineering turns out to be a fertile ground where many software development and maintenance tasks could be formulated as learning problems and approached in terms of learning algorithms.'' 
\end{quote}
	
However, there is a lack of approaches to compare how software engineers fare against machine-learning algorithms for specific software development tasks. This comparison is critical in order to evaluate which S.E. tasks are better performed by automation and which require human involvement  or human-in-the-loop approaches \cite{holzinger2016towards,holzinger2016interactive}. In practice, because there are no explicit comparisons between the tasks performed by engineers and automated procedures, including machine learning, it is often not clear when to use automation in a specific setting. For example, a Brazilian company acquired a software system to select petroleum exploration models automatically, but the engineers decided they could provide a better solution  manually. However, when there was a comparison of the manual solution with the one provided automatically by the system, it became clear that the automated solution was better. This illustrates that a lack of comparisons makes choosing  a manual or an automated solution or a combined human-in-the-loop approach difficult. 

This paper, contains an empirical study  \cite{easterbrook2008selecting} to compare how software engineers and machine-learning algorithms achieve performance and reuse tasks. The empirical study uses a case study involving the creation of a control structure for an autonomous streetlight application. The approach consists of four steps. First, the problem was solved using machine learning to achieve specific performance and reuse of tasks. Second, we asked software engineers with different domain-knowledge levels to provide a solution to achieve the same tasks. Third, we compared how software engineers compare with machine-learning algorithms when accomplishing the performance and reuse tasks based on criteria such as energy consumption and safety. Finally, the results were analyzed to understand which tasks are better performed by either humans or algorithms so that they can work together more effectively.

Such an understanding is essential in realizing novel human-in-the-loop approaches in which machine-learning  procedures assist software developers in achieving tasks. Such human-in-the-loop approaches, which take into account the strengths and weaknesses of humans and machine-learning algorithms, are fundamental not only to provide a basis for cooperative work in software engineering, but also in other application areas.

This paper is organized as follows: Section 2 presents the empirical study  describing research questions, hypotheses and the objective of the study. Section 3 presents the method selected to collect our empirical data. Sections 4 and 5 present the experimental results. Section 6 presents the threats to the validity of our experiment. Section 7 presents the related work. The paper ends with concluding remarks and suggestions for future work.

\subsection{Motivation}\label{sec:motivation}

The theme of this paper, namely whether artificial intelligence such as machine learning, can benefit software engineering, has been investigated since 1986, when Hebert A Simon published a paper entitled  ``Whether software engineering needs  to be artificially intelligent" \cite{simon1986whether}. In this paper, Simon discussed ``the roles that humans now play versus the roles that could be taken over by artificial intelligence in developing computer systems." Notwithstanding, in 1993, Ian Sommerville raised the following question  \cite{sommerville1993artificial}: ``What of the future - can Artificial Intelligence make a contribution to system engineering?" In this paper \cite{sommerville1993artificial}, Sommervile performed a literature review in applications of artificial intelligence to software engineering, and concluded that:
\begin{quote}
``{\bf the contribution of AI} will be in supporting...activities that are characterized by {\bf solutions to problems which are neither right nor wrong but which are more or less appropriate for a particular situation}...For example, requirements specification and analysis which involves extensive consultation with domain experts and in project management."
\end{quote}

Several papers have since investigated the use of Machine Learning  (ML) \cite{michalski2013machine} in solving different software engineering (SE) tasks  \cite{mostow1985foreword, barstow1987artificial,partridge1988artificial,cheung1991survey, partridge1998artificial, van1998inferring, boetticher2001using,padberg2004using,zhang2000applying,zhang2006machine,zhang2005machine,zhang2008machine,khoshgoftaar2003introduction,zhang2009machine,zhang2002machine,marchetto2005evaluating,bouktif2010novel,radlinski2010survey,zhang2011handling,radlinski2011framework,sack2006building,reformat2007introduction,twala2007applying,challagulla2009high,veras2007comparative,birzniece2010interactive,hanchate2010analysis,xu2006machine,wen2012systematic,rashid2012survey,al2013machine,radlinski2012enhancing,pinel2013savant,novitasari2016optimizing,radlinski2012towards,rongfa2012defect,rana2015machine, wang2015use,challagulla2005empirical, kaminsky2004building,kaminsky2004predict,kaminsky2004better,kutlubay2005machine,ren2003learn, ceylan2006software,kastro2008defect,kutlubay2007two,namin2010bayesian,murphy2009metamorphic, afzal2010genetic,murphy2008using,qiu2010framework,murphy2010automatic, afzal2009search,taghi2007empirical,wang2010empirical,jin2008artificial,ferzund2008automated,maqbool2007bayesian, okutan2012software,cotroneo2013learning,zhang2013value,okutan2014software,agarwal2014feature,abaei2014important,okutan2016novel,mu2012software,cahill2013predicting,rana2014adoption,schulz2013predicting,rashid2016r4,rashid2015improvisation,chhabra2014prediction,spanoudakis2003revising,shin2005modeling,shirabad2011predictive,araujo2016architecture,tourwe2004induced,kramer2000gaps,di2002machine,fu2006automated, fu2010semantic,katasonov2008smart,baresi2014short,zhu2014minson,de2016intelligent,birzniece2010use,alrajeh2006inferring, sharifloo2016learning,zoph2016neural,do2017fiot,jacob2010code, amal2014use}. These investigations include approaches to: i) project management \cite{marchetto2005evaluating,bouktif2010novel,radlinski2010survey,zhang2011handling,radlinski2011framework,sack2006building,reformat2007introduction,twala2007applying,challagulla2009high,veras2007comparative,birzniece2010interactive,hanchate2010analysis,xu2006machine,wen2012systematic,rashid2012survey,al2013machine,radlinski2012enhancing,pinel2013savant,novitasari2016optimizing,radlinski2012towards,rongfa2012defect,rana2015machine, wang2015use}, dealing with problems related to cost, time, quality prediction, and resource management; ii) defect prediction \cite{challagulla2005empirical, kaminsky2004building,kaminsky2004predict,kaminsky2004better,kutlubay2005machine,ren2003learn, ceylan2006software,kastro2008defect,kutlubay2007two,namin2010bayesian,murphy2009metamorphic, afzal2010genetic,murphy2008using,qiu2010framework,murphy2010automatic, afzal2009search,taghi2007empirical,wang2010empirical,jin2008artificial,ferzund2008automated,maqbool2007bayesian, okutan2012software,cotroneo2013learning,zhang2013value,okutan2014software,agarwal2014feature,abaei2014important,okutan2016novel,mu2012software,cahill2013predicting,rana2014adoption,schulz2013predicting,rashid2016r4,rashid2015improvisation,chhabra2014prediction}; iii) requirements management, focusing on problems of classifying or representing requirements  \cite{spanoudakis2003revising, shin2005modeling,shirabad2011predictive,araujo2016architecture}, or generating requirements \cite{tourwe2004induced}; iv) software development, such as code generation \cite{kramer2000gaps,di2002machine,fu2006automated, fu2010semantic, jin2008artificial,katasonov2008smart,baresi2014short,zhu2014minson,de2016intelligent}, synthesis \cite{birzniece2010use,alrajeh2006inferring, sharifloo2016learning,zoph2016neural,do2017fiot}, and code evaluation \cite{jacob2010code, amal2014use}.


Most of these papers present successful applications of machine learning in software engineering, showing that ML techniques can provide correct automatic solutions to some SE problems. However,  {\bf very few papers discuss whether or not a domain expert could propose a manual solution more appropriate for the particular situation}. ``More appropriate", means a solution that provides better performance or increases another quality that is important to a particular application scenario, such as user preference  \cite{peng2012user}. For example, in the medical and aviation engineering fields, trust  \cite{abbass2016trusted} in a solution provided to the end-user is an important factor to consider for a solution to be more appropriate. However, although  many  authors \cite{baxt1991use, mazurowski2008training, doiot, morejon2017generating} have been promoting the use of neural networks \cite{haykin1994neural} in medicine, Abbas et al.   \cite{abbass2016trusted} and Castelvecchi \cite{castelvecchi2016can} are among the few authors who questioned: ``what is the trustworthiness of a prediction made by an artificial neural network?"
 

In other application scenarios, such as many of those related to the Internet of Things (IoT) \cite{atzori2010internet,salahuddin2017softwarization}, numerous authors  \cite{katasonov2008smart, zhu2014minson, ayala2015software, do2017fiot} consider the reuse of a solution as an important quality. They agree that to achieve the goal of billions of things connected to the Internet over the next few years \cite{atzori2010internet}, it is necessary to find ways to reduce time to market. For example, it is desirable that the solution or parts of the solution to design autonomous streetlights \cite{de2016intelligent} for a specific scenario could be reused to design streetlights for another scenario.

In particular, the Internet of Things has considerably increased the number of approaches that propose the use of machine learning to automate software development  \cite{katasonov2008smart,baresi2014short,zhu2014minson,de2016intelligent,briot2016multi,do2017fiot}. None of this research contains a comparison of their results to experiments designed by IoT experts. For example, do Nascimento and Lucena \cite{do2017fiot, nathalia:mestrado:15} developed a hybrid framework that uses learning-based and manual program synthesis for the Internet of Things (FIoT). They generated four instances of the framework \cite{nascimento2015modeling, doiot, do2017fiot, nascimento2017engineering} and used learning techniques to synthesize the control structure automatically. These authors stated that the use of machine learning made feasible the development of these applications. However, they did not present any experiment without using learning techniques. In contrast, most of the solutions released for the Internet of Things, such as Apple's HomeKit's approach \cite{homekit} and Samsung Smart Things,  \cite{smartthingssamsung} consider a software developer synthesizing the control structure for each thing manually.

\subsection{Objective}

\begin{figure*}[htb!]
	\centering
	\includegraphics[scale=0.52]{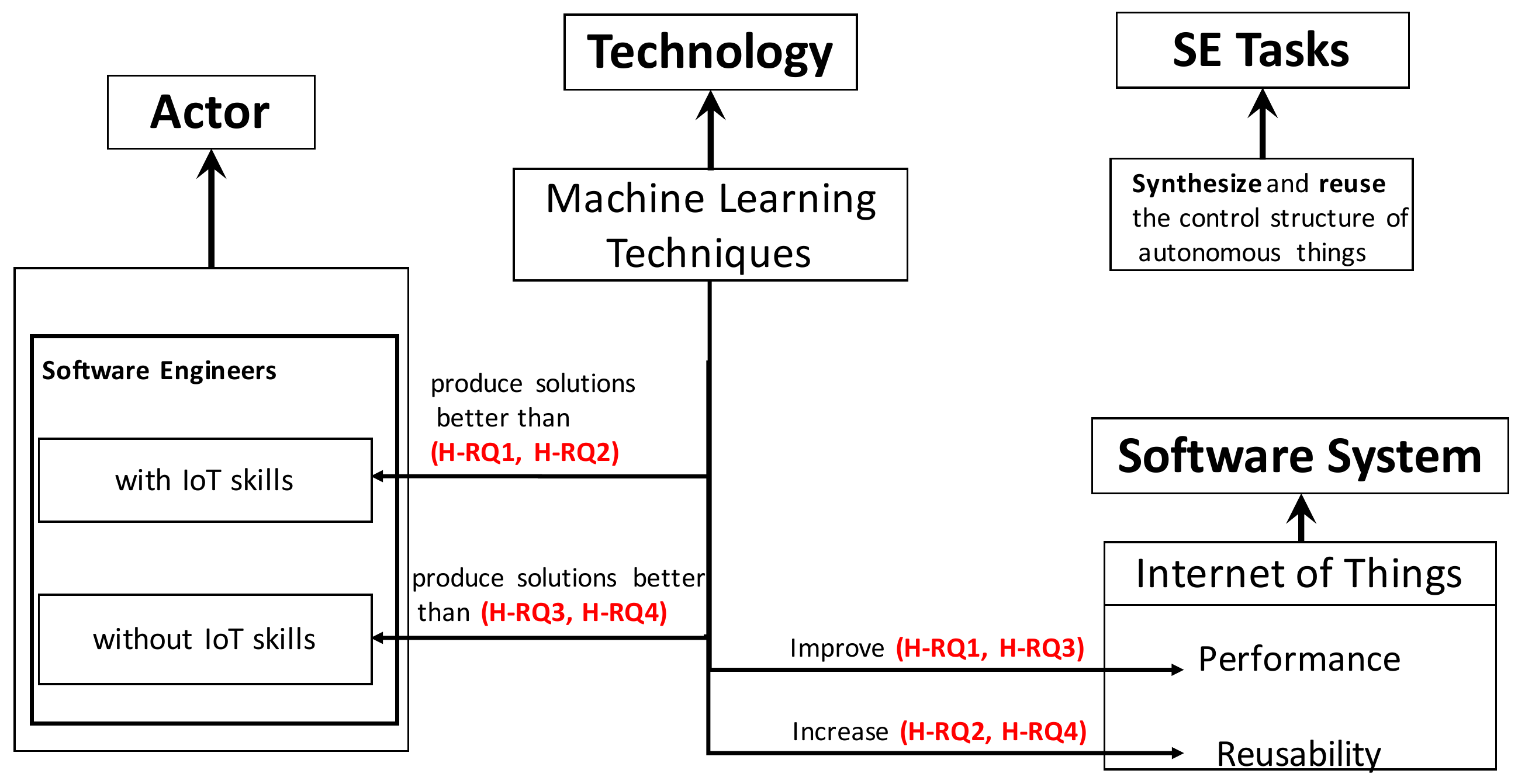} 
	\centering
	\caption{Theory \cite{sjoberg2008building}: Machine Learning can create solutions more appropriate than software engineers in the context of the Internet of Things.}
	\label{fig:theory}
\end{figure*}

In this context, we decided to ask the following question: ``How do software engineers compare with machine-learning algorithms?" To explore this question, we selected the Internet of Things as our application domain and then, compared a solution provided by a skilled IoT professional with a solution provided by a learning algorithm with respect to performance and reuse tasks. In short, Figure \ref{fig:theory} depicts the theory \cite{sjoberg2008building} that we investigate in  this  paper. According to the theory, the variables that we intend to isolate and measure are the performance and reusability achieved from three kinds of solutions: i) solutions provided by learning techniques; ii) solutions provided by software engineers with IoT skills; and iii) solutions provided by software engineers without IoT skills.

To evaluate the relationship among these variables, we performed an empirical study, using FIoT \cite{do2017fiot}. As shown in Figure \ref{fig:theory}, we raised four research questions (RQx) to investigate our theory's propositions (e.g hypotheses (H-RQx)). We present these questions and hypotheses in Section  \ref{sec:empirical}. To collect and analyze our empirical data, we performed a controlled experiment. To perform this experiment, we reproduced the problem of synthesizing the control structure of autonomous streetlights using neuroevolution (i.e. ``a learning algorithm which uses genetic algorithms to train neural networks" \cite{whiteson2005evolving}) presented in  \cite{nascimento2017engineering}. Then, we invited 14 software engineers to provide a solution for the same problem using the same architecture and environment. Lastly, we compared the solution provided by the learning algorithm against the solutions provided by the software engineers. {\bf In this application of autonomous streetlights, we are considering a ``more appropriate" solution as one that presents a better performance in the main scenario \cite{nascimento2017engineering} or can be satisfactorily reused in a new scenario, based on criteria such as minimal energy consumption and safety (that is, maximum visual comfort in illuminated areas)}.

\section{How do software engineers compare with machine-learning algorithms? An Empirical Study Addressing Performance and Reuse in the IoT Domain} \label{sec:empirical}

The experimental goal, based on the Goal-Question-Metric (GQM) method \cite{wohlin2012experimentation} is to use {\bf the Framework for the Internet of Things (FIoT)} for the purpose of {\bf comparing the use of an automated approach against a manual approach when synthesizing the control of autonomous things} with respect to their {\bf performance} and {\bf reuse}. 

For this purpose, we asked four research questions (RQs) and performed a controlled experiment \cite{easterbrook2008selecting} (section \ref{sec:experiment}) to investigate them.


\subsection{Questions}
In terms of synthesizing the control structure of autonomous things, how does the result from a machine learning-based solution differ from solutions provided by...

\begin{quote}
	
%
	
	
	{\bf RQ1}. ...{\bf software engineers with IoT skills} with respect to their {\bf performance}?
	
	{\bf RQ2}. ...{\bf software engineers with IoT skills} with respect to their {\bf re-usability}?
	
	{\bf RQ3}. ...{\bf software engineers without IoT skills} with respect to their {\bf performance}?
	
	{\bf RQ4}. ...{\bf software engineers without IoT skills} with respect to their {\bf re-usability}?
	
\end{quote}


\subsection{Hypotheses} \label{sub:hyp}
Each RQ is based on one or more   hypotheses, which are described next.

%

{\bf H - RQ1}.
\begin{itemize}
	\item H0. An ML-based approach does not improve the performance of autonomous things compared to solutions provided by IoT expert software engineers.
	\item HA. An ML-based approach improves the performance of autonomous things compared to solutions provided by IoT expert software engineers.
\end{itemize}

{\bf H - RQ2}.
\begin{itemize}
	\item H0. An ML-based approach does not increase the reuse of autonomous things compared to solutions provided by IoT expert software engineers.
	\item HA. An ML-based approach increases the reuse of autonomous things compared to solutions provided by IoT expert software engineers.
\end{itemize}

{\bf H - RQ3}.
\begin{itemize}
	\item H0. An ML-based approach does not improve the performance of autonomous things compared to solutions provided by software engineers without experience in IoT development.
	\item HA. An ML-based approach improves the performance of autonomous things  compared to solutions provided by software engineers without experience in IoT development.
\end{itemize}

{\bf H - RQ4}.
\begin{itemize}
	\item H0. An ML-based approach does not increase the reuse of autonomous things compared to solutions provided by software engineers without experience in IoT development.
	\item HA. An ML-based approach increases the reuse of autonomous things compared to solutions provided by software engineers without experience in IoT development.
\end{itemize}

\subsection{The object of the study: The Framework for the Internet of Things (FIoT)} \label{sub:FIoT}
The Framework for the Internet of Things (FIoT) \cite{do2017fiot} is a hybrid software framework that allows the developer to generate controller structures for autonomous things through learning or procedural algorithms. 

If a researcher develops an application using FIoT, the application will contain a Java software component already equipped with modules for detecting autonomous things in an environment, assigning a controller to a specific thing, creating software agents, collecting data from devices and supporting the communication structure among agents and devices.

Some features are variable and may be selected/developed according to the application type, as follows: (i) a control module such as ``if-else", neural network or finite state machine; (2) an adaptive technique to synthesize the controller at design-time, such as reinforcement learning  \cite{sutton1998reinforcement} or genetic algorithm; and (iii) an evaluation process to evaluate the behavior of autonomous things that are making decisions based on the controller.

%

\begin{table}[htb!]
	\centering
	\caption{Implementing FIoT flexible points to synthesize streetlight controllers using a ML-based approach.}
	\begin{tabular}{|l|l|}
		\hline
		\textbf{FIoT Framework}    & \textbf{Light Control Application}                                                                                                                                                                                                                                  \\ \hline
		Controller            & Three Layer Neural Network                                                                                                                                                                                                                            \\ \hline
		Making Evaluation     & \begin{tabular}[c]{@{}l@{}}Collective Fitness Evaluation: \\the solution is evaluated \\based on the energy \\consumption, the number of \\people that finished their \\routes after the \\simulation ends, and the \\total time spent by people \\to move during their trip\end{tabular}  \\ \hline
		\begin{tabular}[c]{@{}l@{}}
			Controller Adaptation \\ at design time
		\end{tabular} 
		& \begin{tabular}[c]{@{}l@{}}Evolutionary Algorithm: \\Generate a pool of \\candidates  to represent the\\neural network parameters\end{tabular}                                                                                                                  \\ \hline
	\end{tabular}
	\label{table:case2}
\end{table}

For example, Table~\ref{table:case2} summarizes how the ``Streetlight Control" application will adhere to the proposed framework using a machine learning-based approach, while extending the FIoT flexible points.

Table~\ref{table:case3} summarizes how the ``Streetlight Control" application will adhere to the proposed framework using a solution provided by a software engineer, while extending the FIoT flexible points.

\begin{table}[htb!]
	\centering
	\caption{Implementing FIoT flexible points to synthesize streetlight controllers using a solution provided by a Software Engineer}
	\begin{tabular}{|l|l|}
		\hline
		\textbf{FIoT Framework}    & \textbf{Light Control Application}                                                                                                                                                                                                                                  \\ \hline
		Controller            & \begin{tabular}[c]{@{}l@{}}if-else module provided by a \\software engineer \end{tabular}                                                                                                                                                                                                                      \\ \hline
		Making Evaluation     & \begin{tabular}[c]{@{}l@{}}Collective Fitness Evaluation: \\the solution is evaluated \\based on the energy \\consumption, the number of \\people that finished their \\routes after the \\simulation ends, and the \\total time spent by people \\to move during their trip\end{tabular} \\ \hline
		\begin{tabular}[c]{@{}l@{}}
			Controller Adaptation \\ at design time
		\end{tabular}  & \begin{tabular}[c]{@{}l@{}}None\end{tabular}                                                                                                                  \\ \hline
	\end{tabular}
	\label{table:case3}
\end{table}

Our goal is to provide both solutions to the same application and compare the results based on the same evaluation process.

\section{Controlled Experiment} \label{sec:experiment}

The first step of the experiment was to reproduce the experiment presented in \cite{nascimento2017engineering} by using a not supervised learning approach. Then, we invited 14 software engineers to provide a solution for the same problem. Finally, we compared the solution provided through the learning algorithm against solutions provided by the participants.

\subsection{Participant Analysis}

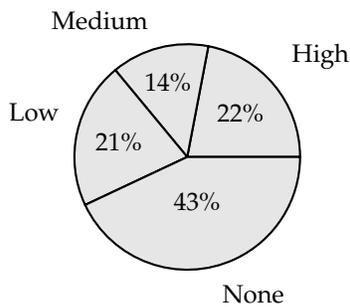
\begin{figure}[!htb]
	\begin{tikzpicture}[scale=1.5]
	
	\newcounter{a}
	\newcounter{b}
	\foreach \p/\t in {22/High, 14/Medium, 21/Low,
		43/None}
	{
		\setcounter{a}{\value{b}}
		\addtocounter{b}{\p}
		\slice{\thea/100*360}
		{\theb/100*360}
		{\p\%}{\t}
	}
	
	\end{tikzpicture}
	\caption{Experience of participants in developing applications based on the Internet of Things.}
	\label{fig:knowledgeiniot}
\end{figure}

As we have described previously, the knowl- edge in the application domain is an important variable in our empirical study. Therefore, before performing the controlled experiment, we asked participants to describe their experience with the development of applications based on the Internet of Things, that is, developing distributed systems with embedded characteristics, such as providing each element of the system with sensors and actuators. As shown in Figure \ref{fig:knowledgeiniot},  43\% of participants have never developed an application based on the Internet of Things and 57\%  have developed at least one application.

\subsection{Experiment: Streetlight Application}
\begin{figure}[!htb]
	\centering
	\includegraphics[width=7.2cm]{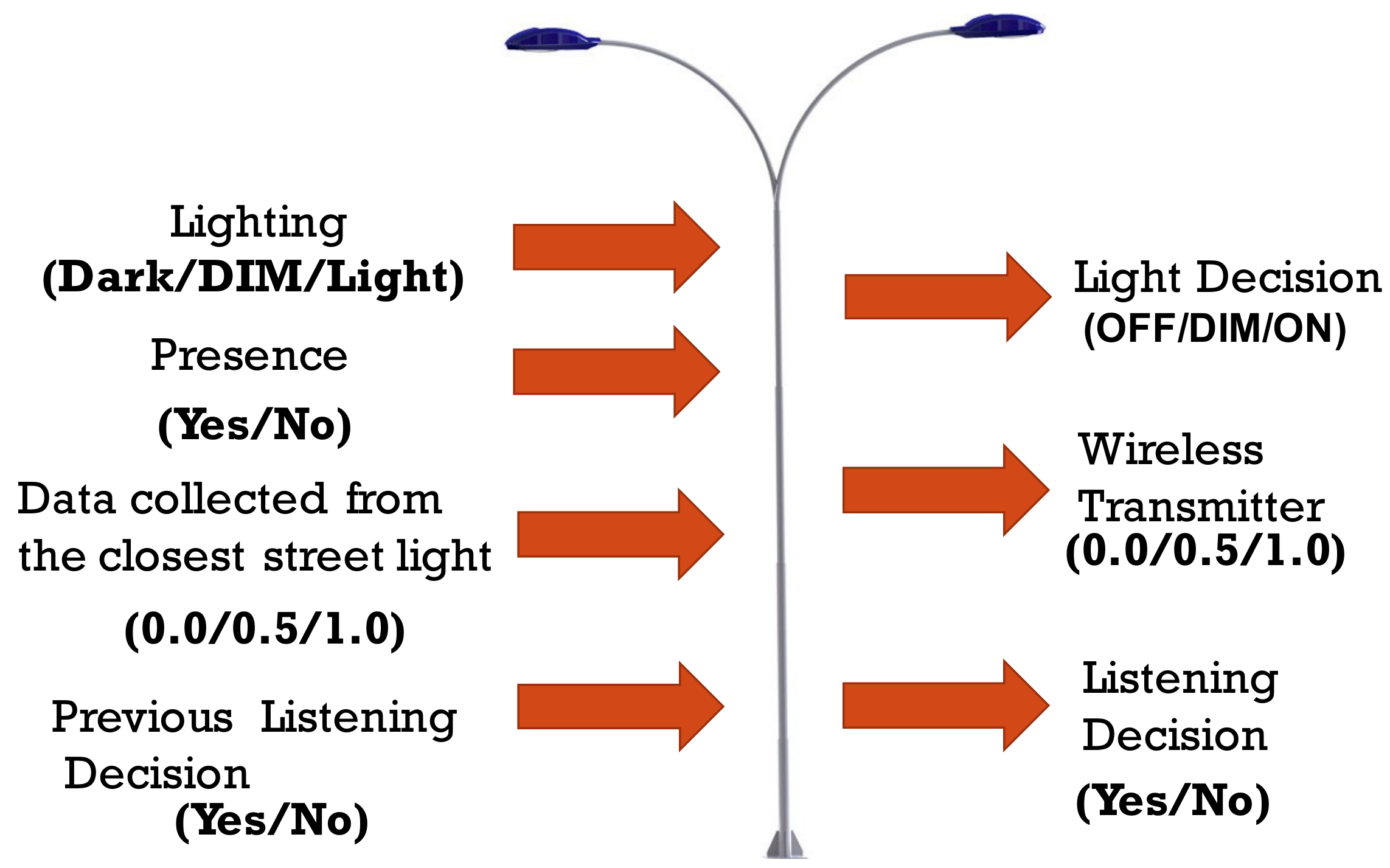}
	\caption{Variables collected and set by streetlights.}
	\label{figure:streetlight}
\end{figure}

In short, our experiment involves developing autonomous streetlights. The overall goal of this application is to reduce the energy consumption while maintaining appropriate visibility in illuminated areas \cite{nascimento2017engineering}. For this purpose, we provided each streetlight with ambient brightness and motion sensors, and an actuator to control light intensity. In addition, we also provided streetlights with wireless communicators as shown in Figure \ref{figure:streetlight}. Therefore, the streetlights are able to cooperate with each other to establish the most likely  routes for passers-by and thus achieve the goal of minimizing energy consumption.

Each streetlight in the simulation has a microcontroller that is used to detect the proximity of a person, and control the closest streetlight. A streetlight can change the status of its light to  ON, OFF or DIM.

{\bf Each streetlight has to execute three tasks every second: data collection, decision-making and action enforcement}. The first task consists of receiving data related to people flow, ambient brightness, data from the neighboring streetlights and current light status (activation level  of sensors and the previous output value of listeningDecision). The second task consists of analyzing collected data and making decisions about actions to be enforced.  The last task is the action enforcement, which consists of setting the value of three output variables: (i) listeningDecision, that enables the streetlight to receive signals from neighboring streetlights in the next cycle; (ii) wirelessTransmitter, a signal value to be transmitted to neighboring streetlights; and (iii) lightDecision, that activates the light's OFF/DIM/ON functions.

The interested reader may consult a more extensive paper about the application scenario \cite{nascimento2017engineering} \footnote{All documents that we prepared to explain this application scenario to participants are available at \\
	\href{http://www.inf.puc-rio.br/~nnascimento/projects.html}{http://www.inf.puc-rio.br/\~ .nnascimento/projects.html}}. 

\subsubsection{The Challenge}

\begin{figure}[!htb]
	\centering
	\includegraphics[width=7.2cm]{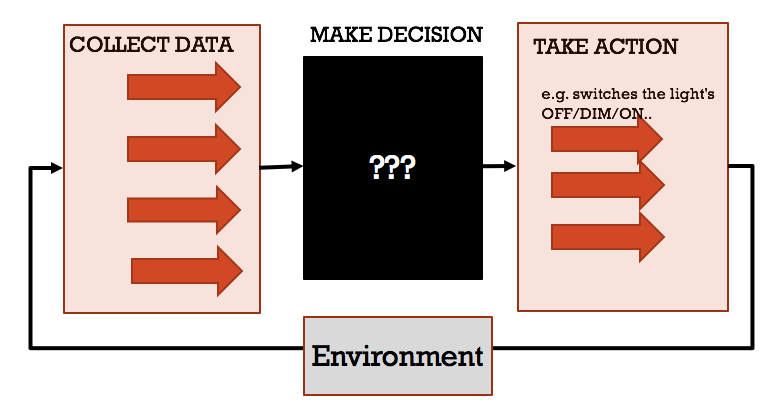}
	\caption{The challenge: how does a streetlight make decisions based on collected data?}
	\label{figure:cycle}
\end{figure}

As we explained to the participants, the tasks of collecting data and enforcing actions have already been implemented. The challenge was to provide a solution for the task of making decisions, as depicted in Figure  \ref{figure:cycle}.

\begin{figure*}[!htb]
	\centering
	\includegraphics[width=11.2cm]{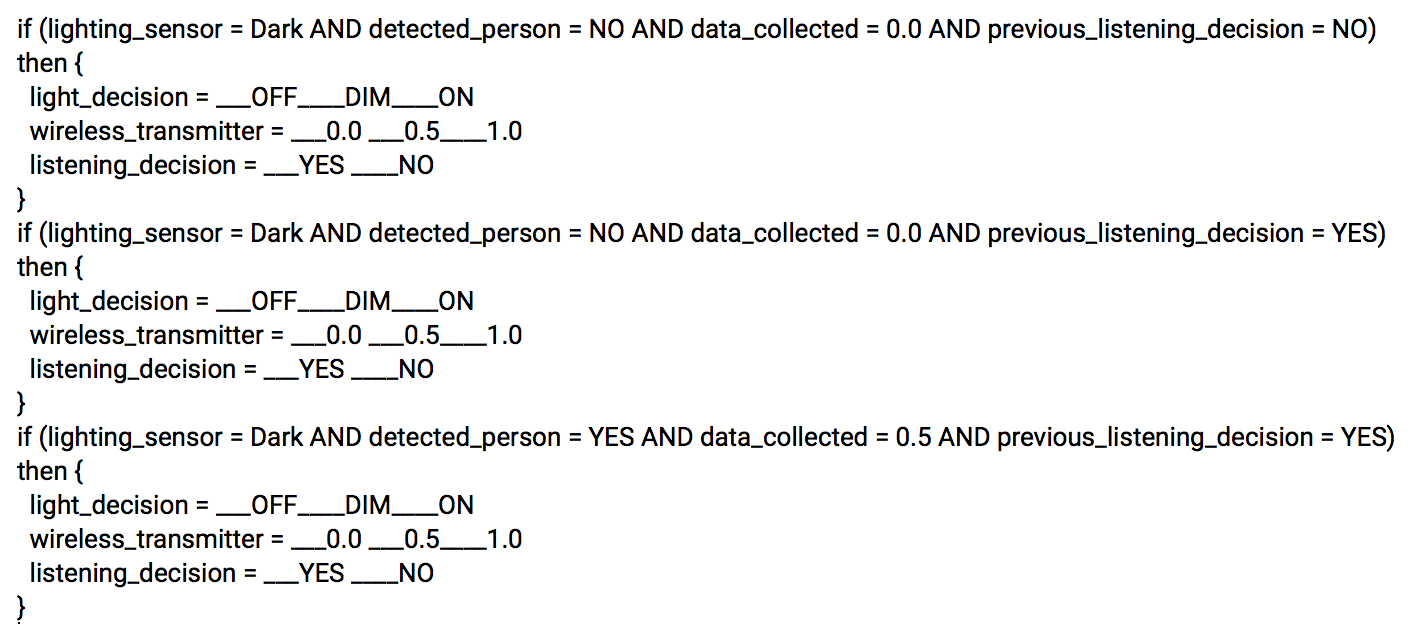}
	\caption{Small portion of the pseudocode of the decision module that was filled by participants. }
	\label{figure:pseudocode}
\end{figure*}

We provided pseudocode that considered all possible combinations of input variables. Then, participants decided how to set output variables according to the collected data. Part \footnote{The pseudocode that we provided to participants is available at:\\
	\href{http://www.inf.puc-rio.br/~nnascimento/projects.html}{http://www.inf.puc-rio.br/\~ .nnascimento/projects.html}} of this pseudocode is depicted in Figure \ref{figure:pseudocode}. 

\begin{figure*}[!htb]
	\centering
	\includegraphics[width=11.2cm]{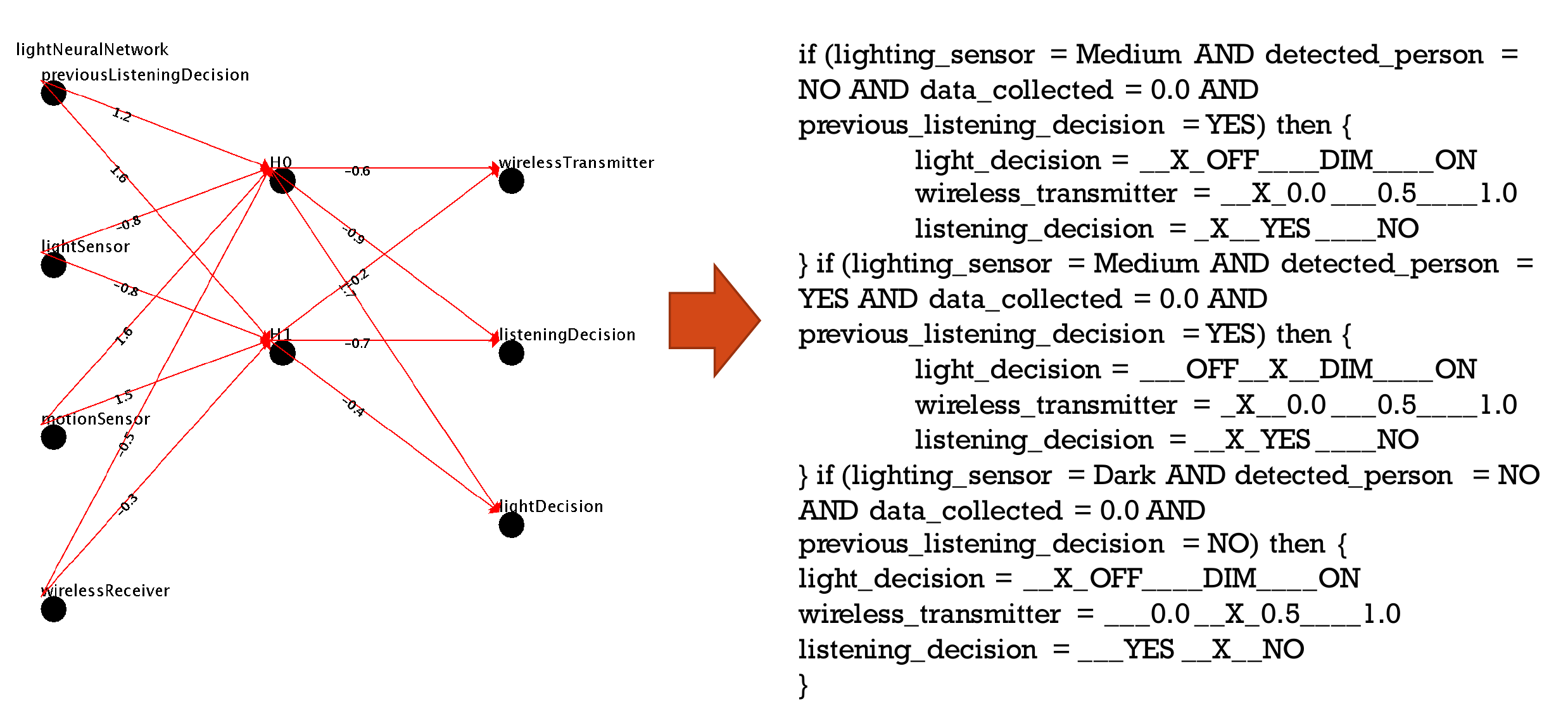}
	\caption{Small portion of the rule decisions that was synthesized according to the learning-based approach. }
	\label{figure:learning}
\end{figure*}

Each participant provided a different solution. Therefore, we conducted the experiment by using each one. In addition, we also considered a ``zeroed" solution, which always sets all values to zero. This zeroed solution is supposed to be the worst solution, since streetlights will always switch their lights to OFF.

\subsubsection{The solution generated by a machine-learning algorithm}

We compared the results from all of these approaches to the result produced using the machine learning approach. As do Nascimento and Lucena explain in  \cite{nascimento2017engineering}, the learning approach uses a three-layer feedforward neural network combined with an evolutionary algorithm to generate decision rules automatically.  Figure \ref{figure:learning} depicts some of the rules that were generated by the evolved neural network. The interested reader can 
consult more extensive papers  \cite{nascimento2017engineering, do2017fiot} or read Nascimento's dissertation \cite{nathalia:mestrado:15} (chap. ii, sec. iii).

Based on the generated rules and the system execution, we observe that using the solution provided by the neural network, only the streetlights with broken lamps emit ``0.5" from their wireless transmitters.      

In addition, we also observed that a streetlight that is not broken switches its lamp ON if it detects a person’s proximity or receives ``0.5" from a wireless transmitter.


\subsubsection{Scenario constraints}

Before starting a solution,  each participant should consider the following constraints:
\begin{itemize}
	\item Do not take light numbering into account, since your solution may be used in different scenarios (see an example of a scenario in Figure  \ref{figure:simulation}).
	\item Three streetlights will go dark during the simulation.
	\item People walk along different paths starting at random departure points. Their role is to complete their routes by reaching a destination point. The number of people that finished their routes after the simulation ends, and the total time spent by people moving during their trip are the most important factors for a good solution.
	\item A person can only move if his current and next positions are not completely dark. In addition, we also consider that people walk slowly if the place is partially devoid of light.
	\item The energy consumption also influences  the solution evaluation.
	\item The energy consumption is proportional to the light status (OFF/DIM/ON).
	\item We also consider the use of the wireless transmitter to calculate energy consumption (if the streetlight emits something different from ``0.0", it consumes 0.1 of energy).

\end{itemize}

Therefore, each solution is evaluated after the simulation ends based on the energy consumption, the number of people that finished their routes after the simulation ends, and the total time spent by people moving during their trip.

\begin{equation}
pPeople = \frac{(completedPeople \times 100)}{totalPeople}
\label{eq:percentPeople}
\end{equation}
\begin{equation}
pEnergy = \frac{(totalEnergy \times 100)}{(\frac{11 \times (timeSimulation \times totalSmartLights)}{10})}	
\label{eq:percentEnergy}
\end{equation}
\begin{equation}
pTrip =\frac{(totalTimeTrip \times 100)}{((\frac{3 \times timeSimulation}{(2)}) \times totalPeople)}
\label{eq:percentTrip}
\end{equation}
\begin{equation} 
\begin{split}
fitness = (1.0 \times pPeople) - (0.6 \times pTrip) -\\ 
(0.4 \times pEnergy)
\end{split}
\label{eq:fitness}
\end{equation}

Equations (1) through (4) show the values to be calculated for the evaluation in which \begin{math} pPeople \end{math} is the percentage of the number of people that completed their routes before the end of the simulation out of the total number of people in the simulation; \begin{math} pEnergy \end{math} is the percentage of energy that was consumed by streetlights out of the maximum energy value that could be consumed during the simulation. We also considered the use of the wireless transmitter to calculate energy consumption;  \begin{math} pTrip \end{math} is the percentage of the total duration time of people’s trips out of the maximum time value that their trip could spend; and  \begin{math} fitness \end{math} is the fitness of each candidate that encodes the proposed solution.

\subsubsection{Example - Simulating the environment}

\begin{figure}[!htb]
	\centering
		\includegraphics[height=1.9in, width=2.85in]{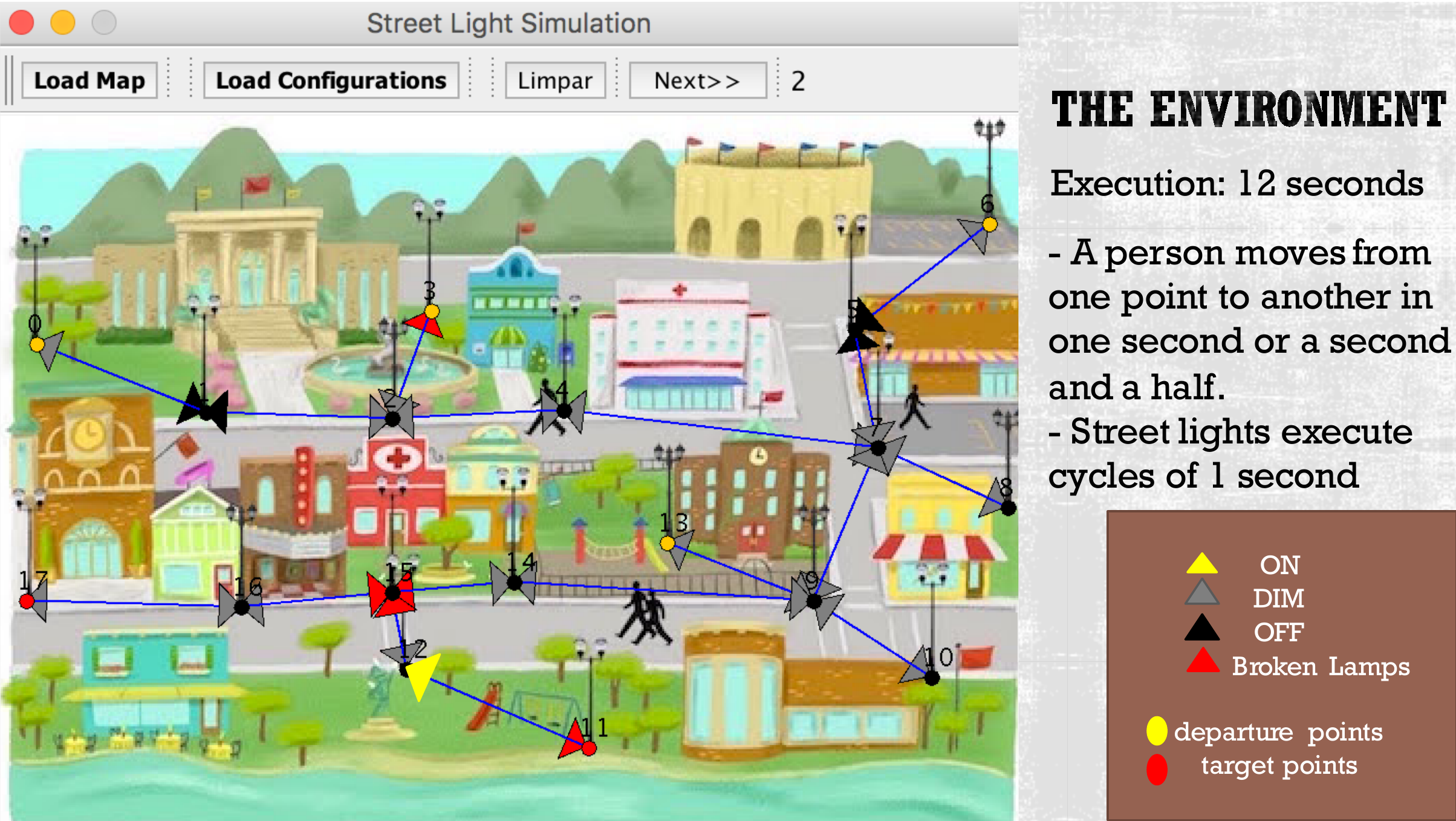}
	\caption{Simulated Neighborhood.}
	\label{figure:simulation}
\end{figure}

We showed participants the same simulated neighborhood scenario that was used by the genetic algorithm to evolve the neural network.  Figure \ref{figure:simulation} depicts the elements that are part of the application namely, streetlights, people, nodes and edges.


Nascimento and Lucena \cite{nascimento2017engineering} modeled the scenario as a graph, in which a node represents a streetlight position and an edge represents the smallest distance between two streetlights. The graph representing the streetlight network consists of 18 nodes and 34 edges. Each node represents a streetlight. In the graph, the yellow, gray, black and red triangles represent the streetlight status (ON/DIM/OFF/Broken Lamp). Each edge is two-way and links two nodes. In addition, each edge has a light intensity parameter that is the sum of the environmental light and the brightness from the streetlights in its nodes. Their goal is to simulate different lighting in different neighborhood areas.

\begin{figure}[!htb]
	\centering
	\includegraphics[height=2in, width=2.75in]{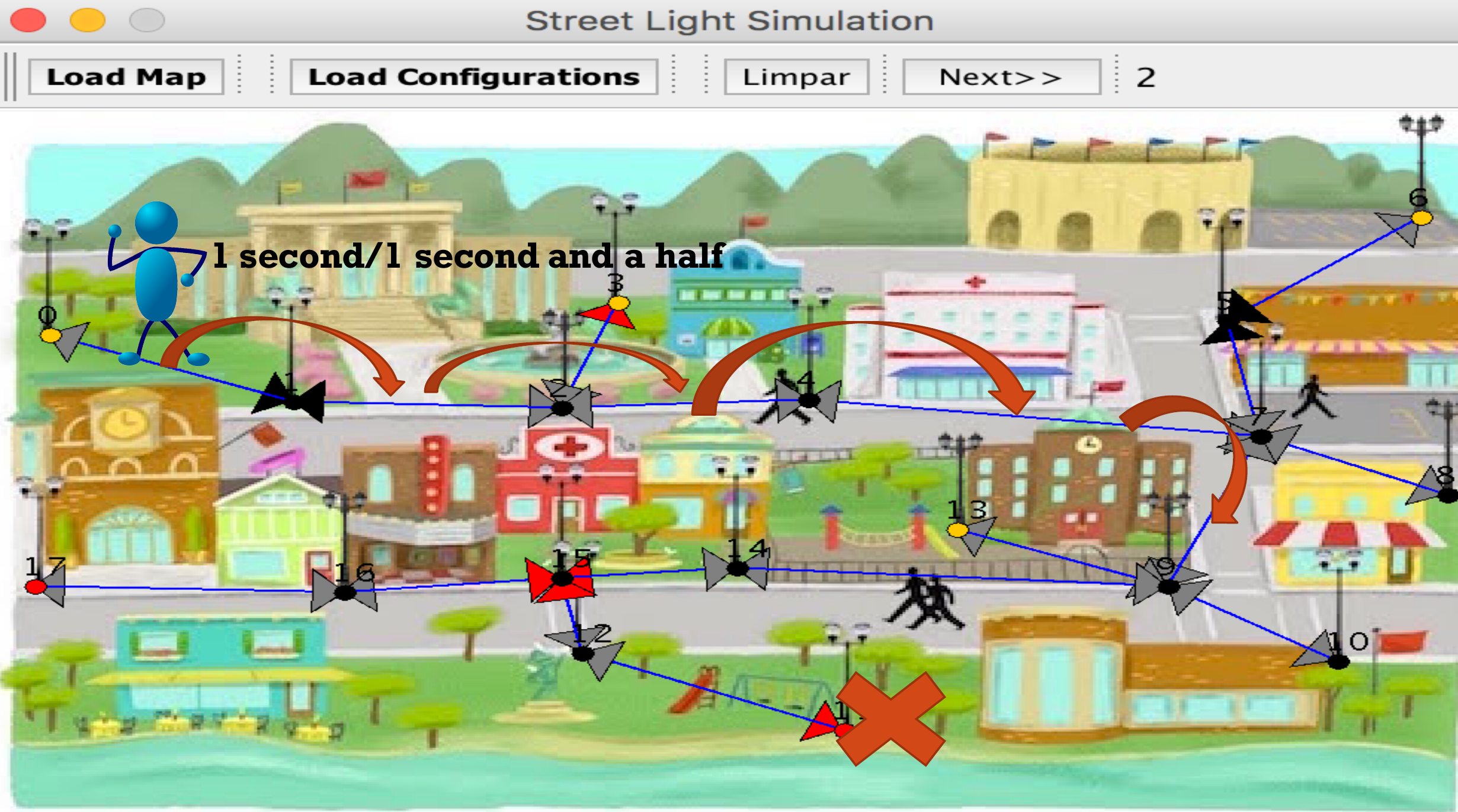}
	\caption{Person moving in the simulated Neighborhood.}
	\label{figure:simulationperson}
\end{figure}

As depicted in Figure \ref{figure:simulationperson}, only one person was started in the scenario that we showed to participants. For instance, the person starting at point 0 has point 11 as a target. We ask participants to provide a solution to streetlights to assure that this person will conclude his route before the simulation ends after 12 seconds.

\subsubsection{New Scenario: Unknown environment}

\begin{figure}[!htb]
	\centering
	\includegraphics[height=2.6in, width=2.7in]{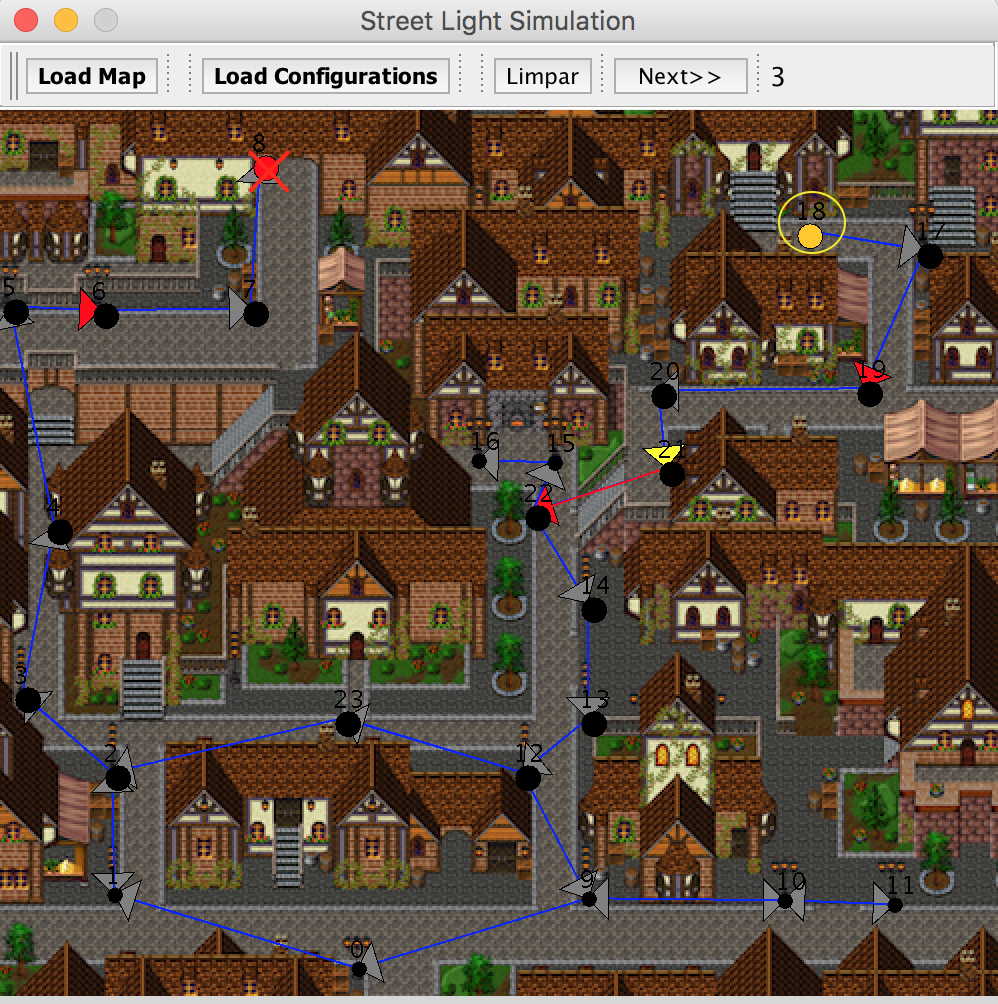}
	\caption{Simulating a new neighborhood.}
	\label{figure:newscenario}
\end{figure}

The second step of the experiment consists of executing solutions from participants and the learning approach in a new scenario, but with the same constraints. This scenario, that is depicted in Figure \ref{figure:newscenario} was not used by the learning algorithm and was not presented to participants.

The goal of this new part of the experiment is to verify if the decision module that was designed to control streetlights in the first scenario can be reused in another scenario.

In this new scenario, we also only started one person, who has the point 18 (yellow point) as departure and the point 8 as target. As the scenario is larger, we established a simulation time of 30 seconds.

\section{Experiment - Part 1 - Results}

We executed the experiment 16 times, only changing the decision solution of the autonomous streetlights. In the first instance, we set all outputs to zero (the zeroed solution) during the whole simulation, which is supposed to be the worst solution. For example, streetlights never switch their lights ON. In the second instance, we executed the experiment using the best solution that was found by the learning algorithm, according to the experiment presented in  \cite{nascimento2017engineering}. Then, we executed the simulation for the solution provided by each one of the 14 participants \footnote{All files that were generated during the development of this work, such as executable files and participants' solutions results, are available at \\
	\href{http://www.inf.puc-rio.br/~nnascimento/projects.html}{http://www.inf.puc-rio.br/\~ .nnascimento/projects.html}}.

To provide a controlled experiment and be able to compare the different solutions, we started with only one person in the scenario and manually we set the parameters that were supposed to be randomly selected, such as departure and target points and broken lamps.

\begin{figure}[!htb]
	\centering
	\includegraphics[width=7.2cm]{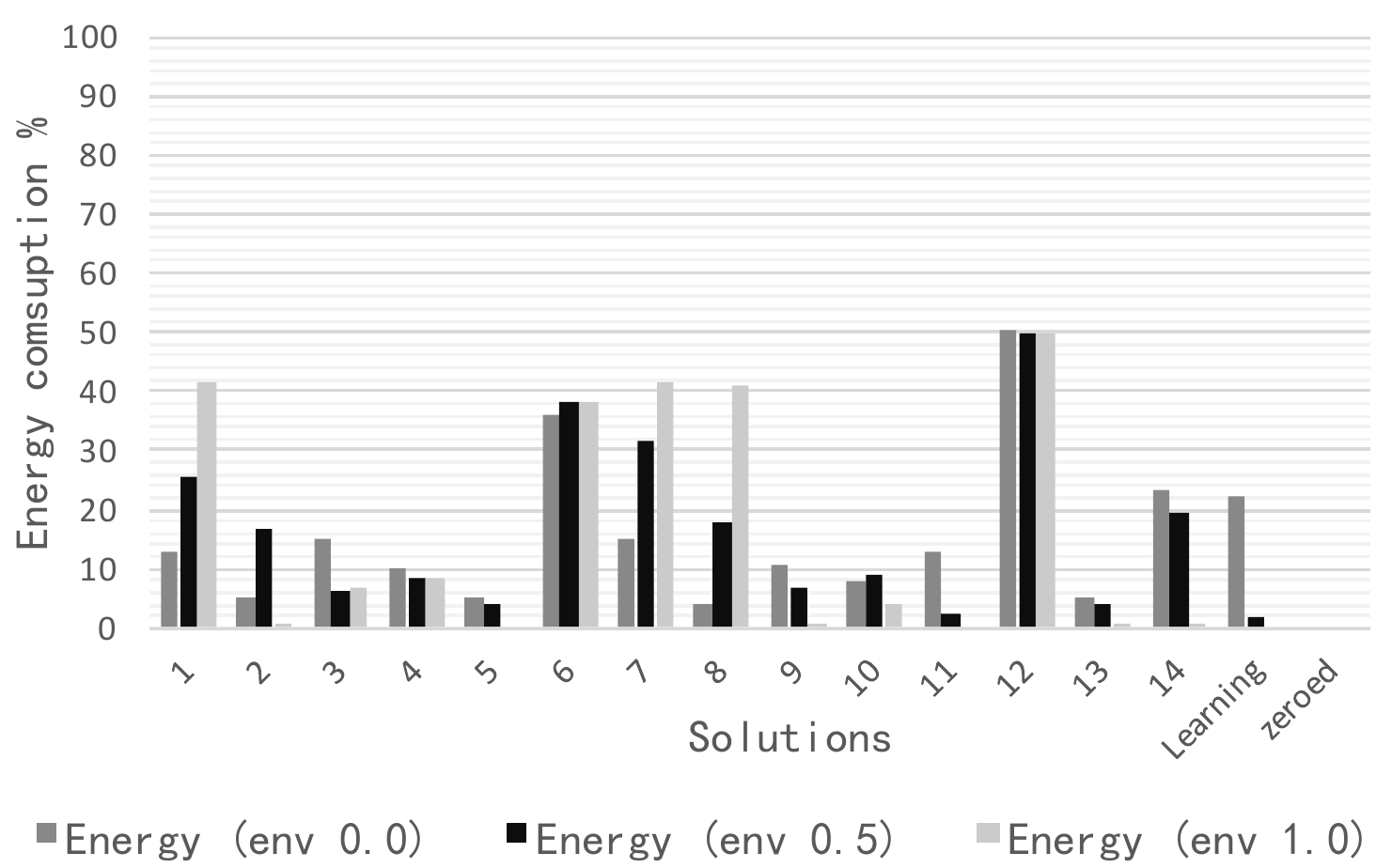}
	\caption{Scenario1: Percentage of energy spent in different parts of the day according to the participant solutions.}
	\label{fig:energy}
\end{figure}

Each experiment execution consists of executing the simulated scenario three times: (i) night (environmental light is equal to 0.0); (ii) late afternoon (environmental light is equal to 0.5); and (iii) morning (environmental light is equal to 1.0). The main idea is to determine how the solution behaves during different parts of the day. Figure \ref{fig:energy} depicts the percentage of energy that was spent according to the environmental light for each one of the 16 different solutions. As we described previously, we also considered the use of the wireless transmitter to calculate energy consumption. As expected, as streetlights using the zeroed decision never switch their lights ON and never emit any signal, the energy consumed using this solution is always zero. It is possible to observe that only the solutions provided by the learning algorithm and by the 5th and 11th participants do not expend energy when the environmental light is maximum.  In fact, according to the proposed scenario, there is no reason to turn ON streetlights during the period of the day with maximum illumination.

\begin{figure}[!htb]
	\centering
	\includegraphics[width=7.2cm]{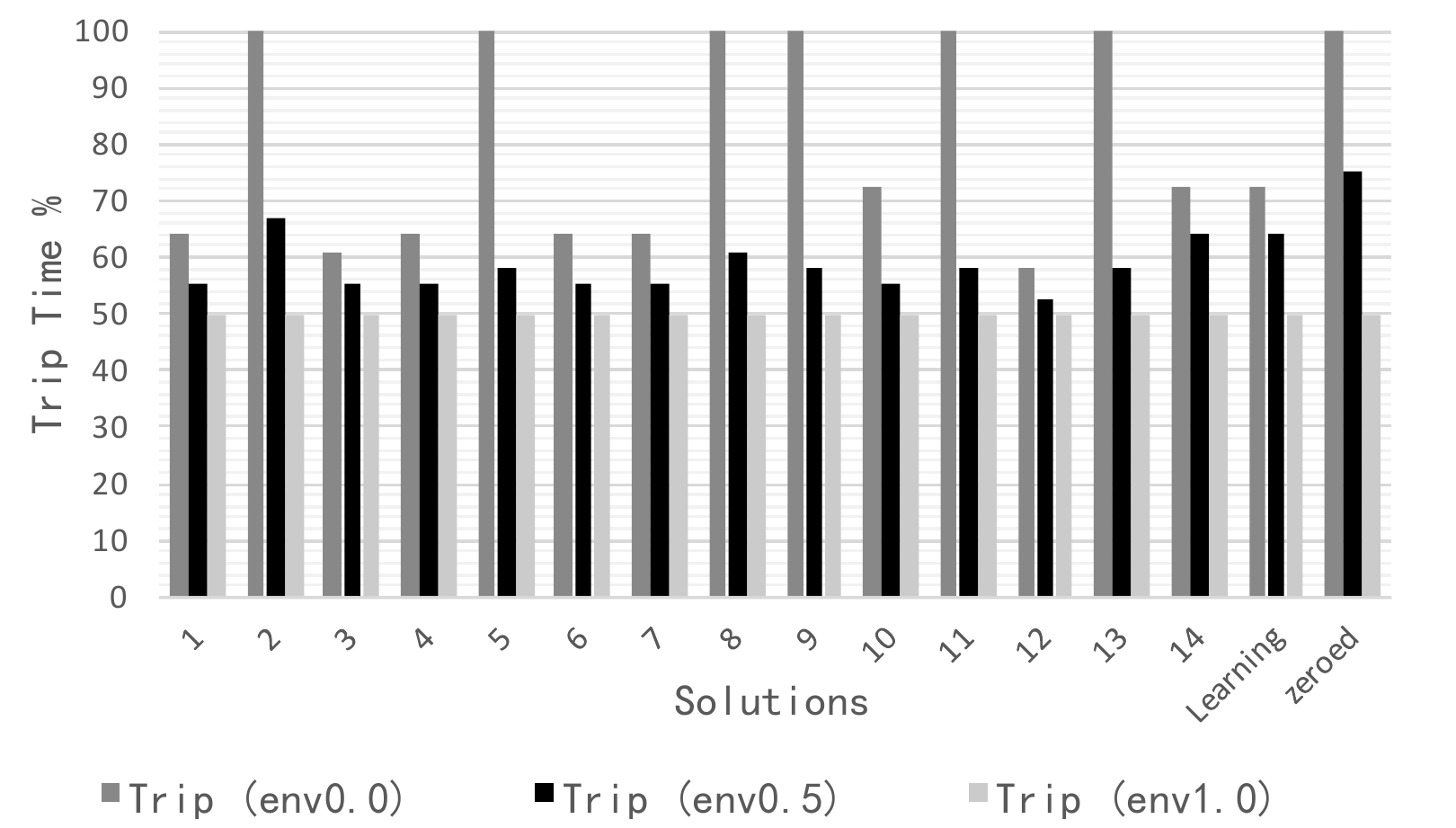}
	\caption{Scenario1: Percentage of time spent by person to conclude his route based on different parts of the day according to the participant solutions.}
	\label{fig:trip}
\end{figure}

Figure \ref{fig:trip} depicts the percentage of time that was spent by the unique person in each one of the simulations. As shown, the higher difference between solutions occurs at night. If the time is 100\%, it means that the person did not complete the route, thus the solution did not work

\begin{figure}[!htb]
	\centering
	\includegraphics[width=7.4cm]{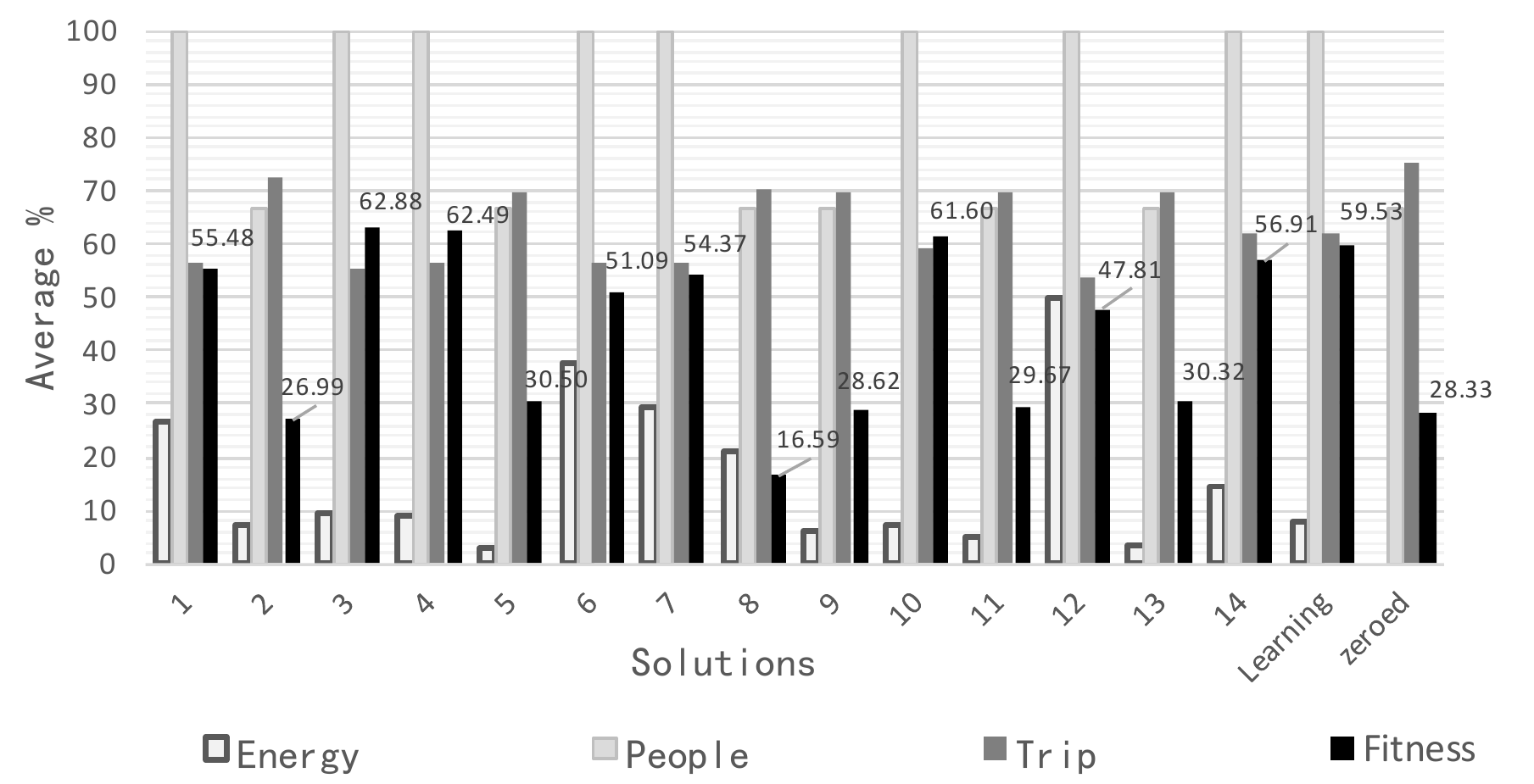}
	\caption{Scenario1: Average of energy, trip and fitness calculated for the different parts of the day according to the participant solutions.}
	\label{fig:average}
\end{figure}

Besides presenting the results of the different solutions in different parts of the day, the best solution must be the one that presents the best result for the whole day. Thus, we calculated the average of each one of the parameters (energy, people, trip and fitness) that was achieved by solutions in different parts of the day. Figure \ref{fig:average} depicts a common average. We also calculated a weighted average, taking into account the duration of the parts of the day (we  considered 12 hours for the night period, 3h for dim and 9h for the morning), but the results were very similar.


As shown in Figure \ref{fig:average}, based on  the fitness average, three participants namely 3, 4 and 10 provided a solution slightly better than the solution provided by the learning algorithm.  Five other participants provided a solution that works and the remaining six  provided a solution that does not work. As explained earlier, we have been considering an incorrect solution as  one in which the person did not finish the route before the simulation ends. Even increasing the simulation time did not allow the person to  finish the  route.

\subsection{Discussion: Participants Knowledge in IoT Versus Results}

After executing the solution proposed by each participant, we connect that solution's results with the participant's knowledge in the IoT domain, as shown in Table  \ref{table:survey}. 


\begin{table}[!htb]
	\centering
	\caption{Correlation between participants expertises in the Internet of Things with their solution results.} 
     \label{table:survey}
	\begin{tabular}{|c|c|c|c|}
		\hline
		\rowcolor[HTML]{000000} 
		{\color[HTML]{FFFFFF} \begin{tabular}[c]{@{}c@{}}Software\\ Engineer\end{tabular}} & {\color[HTML]{FFFFFF} \begin{tabular}[c]{@{}c@{}}Experience \\ with IoT \\ Development\\ (None/Low/\\ Medium/\\ High)\end{tabular}} & {\color[HTML]{FFFFFF} \begin{tabular}[c]{@{}c@{}}Solution\\ Performance\\ (Fitness\\ Average)\end{tabular}} & {\color[HTML]{FFFFFF} \begin{tabular}[c]{@{}c@{}}Does \\ the\\ solution \\ work?\end{tabular}} \\ \hline
		\rowcolor[HTML]{FFFFFF} 
		1                                                                                  & High                                                                                                                                & 55.48                                                                                                       & Y                                                                                              \\ \hline
		\rowcolor[HTML]{FFFFFF} 
		2                                                                                  & None                                                                                                                                & 26.99                                                                                                       & N                                                                                              \\ \hline
		\rowcolor[HTML]{C0C0C0} 
		3                                                                                  & High                                                                                                                                & 62.88                                                                                                       & Y                                                                                              \\ \hline
		\rowcolor[HTML]{C0C0C0} 
		4                                                                                  & Low                                                                                                                                 & 62.49                                                                                                       & Y                                                                                              \\ \hline
		\rowcolor[HTML]{FFFFFF} 
		5                                                                                  & None                                                                                                                                & 30.50                                                                                                       & N                                                                                              \\ \hline
		\rowcolor[HTML]{FFFFFF} 
		6                                                                                  & Low                                                                                                                                 & 51.09                                                                                                       & Y                                                                                              \\ \hline
		\rowcolor[HTML]{FFFFFF} 
		7                                                                                  & Medium                                                                                                                              & 54.37                                                                                                       & Y                                                                                              \\ \hline
		\rowcolor[HTML]{FFFFFF} 
		8                                                                                  & None                                                                                                                                & 16.59                                                                                                       & N                                                                                              \\ \hline
		\rowcolor[HTML]{FFFFFF} 
		9                                                                                  & High                                                                                                                                & 28.62                                                                                                       & N                                                                                              \\ \hline
		\rowcolor[HTML]{C0C0C0} 
		10                                                                                 & None                                                                                                                                & {\color[HTML]{343434} 61.60}                                                                                & Y                                                                                              \\ \hline
		\rowcolor[HTML]{FFFFFF} 
		11                                                                                 & None                                                                                                                                & 29.67                                                                                                       & N                                                                                              \\ \hline
		\rowcolor[HTML]{FFFFFF} 
		12                                                                                 & Medium                                                                                                                              & 47.81                                                                                                       & Y                                                                                              \\ \hline
		\rowcolor[HTML]{FFFFFF} 
		13                                                                                 & None                                                                                                                                & 30.32                                                                                                       & N                                                                                              \\ \hline
		\rowcolor[HTML]{FFFFFF} 
		14                                                                                 & Low                                                                                                                                 & 56.91                                                                                                       & Y                                                                                              \\ \hline
		\rowcolor[HTML]{FFFFFF} 
		\textbf{Learning}                                                                  & \multicolumn{1}{l|}{\cellcolor[HTML]{FFFFFF}{\color[HTML]{FFFFFF} }}                                                                & \textbf{59.53}                                                                                              & \textbf{Y}                                                                                     \\ \cline{1-1} \cline{3-4} 
		\rowcolor[HTML]{FFFFFF} 
		zeroed                                                                             & \multicolumn{1}{l|}{\multirow{-2}{*}{\cellcolor[HTML]{FFFFFF}{\color[HTML]{FFFFFF} \textbf{}}}}                                     & 28.33                                                                                                       & N                                                                                              \\ \cline{1-1} \cline{3-4} 
	\end{tabular}
\end{table}

{\bf We observe a significant difference between results from software engineers with any experience in IoT development and results from software engineers without experience in IoT development}. Participant 10 is the only individual without knowledge of IoT that provided a solution that works and participant 9 is the only individual with any knowledge of IoT that did not provide a working solution.


\subsection{Hypothesis Testing}

In this section, we investigate the hypotheses related to the solutions' performance evaluation (i.e H-RQ1 and H-RQ3), as presented in subsection \ref{sub:hyp}. Thus, we performed {\bf statistical analyses}, as described by Peck and Devore  \cite{peck2011statistics}, of the measures presented in Table  \ref{table:survey}. 

As shown in Table \ref{table:average}, we separated the results of the experiments into two groups: i) software engineers with IoT knowledge and ii) software engineers without IoT knowledge. Then, we calculated the mean and the standard deviation of the results achieved by each group performing the experiment and compare each result against the value achieved using the ML-based solution. 

\begin{table*}[!htb]
	\centering
	\caption{Data to perform test statistic.}
	\label{table:average}
	\begin{tabular}{|c|c|c|clccc}
		\hline
		\rowcolor[HTML]{C0C0C0} 
		Variable                                                                                 & \begin{tabular}[c]{@{}c@{}}n\\ samples\end{tabular} & \begin{tabular}[c]{@{}c@{}}Highest\\ value\end{tabular} & \multicolumn{1}{c|}{\cellcolor[HTML]{C0C0C0}\begin{tabular}[c]{@{}c@{}}Mean\\ \begin{math} \overline{x} \end{math}\end{tabular}} & \multicolumn{1}{l|}{\cellcolor[HTML]{C0C0C0}Median} & \multicolumn{1}{c|}{\cellcolor[HTML]{C0C0C0}\begin{tabular}[c]{@{}c@{}}Standard \\ deviation\\ \begin{math} \sigma \end{math}\end{tabular}} & \multicolumn{1}{c|}{\cellcolor[HTML]{C0C0C0}\begin{tabular}[c]{@{}c@{}}Degrees\\ of\\ freedom\\ (n-1)\end{tabular}} & \multicolumn{1}{l|}{\cellcolor[HTML]{C0C0C0}\begin{tabular}[c]{@{}l@{}}t \\ critical\\ value\\ (.99\%)\end{tabular}} \\ \hline
		\begin{tabular}[c]{@{}c@{}}Software \\ Engineers\end{tabular}                            & 14                                                    & 62.88                                                   & \multicolumn{1}{c|}{43.95}                                                                       & \multicolumn{1}{l|}{49.45}                          & \multicolumn{1}{c|}{16.00}                                                                                        & \multicolumn{1}{c|}{13}                                                                                             & \multicolumn{1}{c|}{2.65}                                                                                            \\ \hline
		\begin{tabular}[c]{@{}c@{}}Software \\ Engineers\\ with IoT\\ knowledge\end{tabular}     & 8                                                     & 62.88                                                   & \multicolumn{1}{c|}{52.46}                                                                       & \multicolumn{1}{l|}{54.92}                          & \multicolumn{1}{c|}{10.91}                                                                                        & \multicolumn{1}{c|}{7}                                                                                              & \multicolumn{1}{c|}{3.00}                                                                                            \\ \hline
		\begin{tabular}[c]{@{}c@{}}Software \\ Engineers\\ without IoT \\ knowledge\end{tabular} & 6                                                     & 61.60                                                   & \multicolumn{1}{c|}{32.61}                                                                       & \multicolumn{1}{l|}{30.00}                          & \multicolumn{1}{c|}{15.15}                                                                                        & \multicolumn{1}{c|}{5}                                                                                              & \multicolumn{1}{c|}{3.37}                                                                                            \\ \hline
		\begin{tabular}[c]{@{}c@{}}Machine-\\ learning\\ based \\ approach\end{tabular}          & 1                                                     & 59.53                                                   &                                                                                                  &                                                     &                                                                                                                   &                                                                                                                     & \multicolumn{1}{l}{}                                                                                                 \\ \cline{1-3}
	\end{tabular}
\end{table*}


\subsubsection{How does the evaluation result from a machine learning-based solution differ from solutions provided by IoT expert software engineers with respect to their performance?}

{\bf H - RQ1}.
\begin{itemize}
	\item H0. An ML-based approach does not improve the performance of autonomous things compared to solutions provided by IoT expert software engineers.
	\item HA. An ML-based approach improves the performance of autonomous things compared to solutions provided by IoT expert software engineers.
\end{itemize}

The first null hypothesis H0 claims that there is no difference between the mean performance for IoT expert software engineers' solutions and the ML-based approach solution. The alternative hypothesis claims that the ML-based approach solution improves the performance of the application in comparison to IoT expert software engineers' solutions. Thus, the claim is that the true IoT expert software engineers' solutions mean is below the performance achieved by the ML-based approach, that is = 59.53.

Therefore, we used the ML performance as our \begin{math} hypothesizedvalue  \end{math} to test the following one-sided hypothesis:

H0: \begin{math} \mu _se \end{math}  = 59.53

H1: \begin{math} \mu _se  \end{math}  $< 59.53$

where \begin{math} \mu _se  \end{math} denotes the true mean performance for all IoT expert software engineers' solutions.

For instance, we restricted the population sample to the number of software engineers that confirmed having experience with developing applications for the Internet of Things.  As shown in Table \ref{table:average}, the performance mean (\begin{math} \overline{x} \end{math}) of the IoT expert software engineers' solutions is 52.46 and the standard deviation (\begin{math} \sigma \end{math}) is 10.91. To verify if the data that we have is sufficient to accept the alternative hypothesis, we need to verify the probability of rejecting the null hypothesis  \cite{peck2011statistics}. Assuming that the H0 is true, and using a statistical significance level \cite{peck2011statistics} of 0.01 (the chance of one in 100 of making an error), we computed the test statistic (\begin{math}  t-statistic\end{math}), as follows \cite{peck2011statistics}:

\begin{equation}
t-statistic: t =\frac{(\overline{x} - hypothesized value)}{(\frac{\sigma}{\sqrt {n}} )}
\label{eq:tstatictis}
\end{equation}

\begin{equation}
t-statistic: t =\frac{(52.46 - 59.53)}{(\frac{10.91}{\sqrt {8}} )} = -1.83
\label{eq:test02}
\end{equation}

\begin{figure}[!htb]
	\centering
	\includegraphics[width=7.2cm]{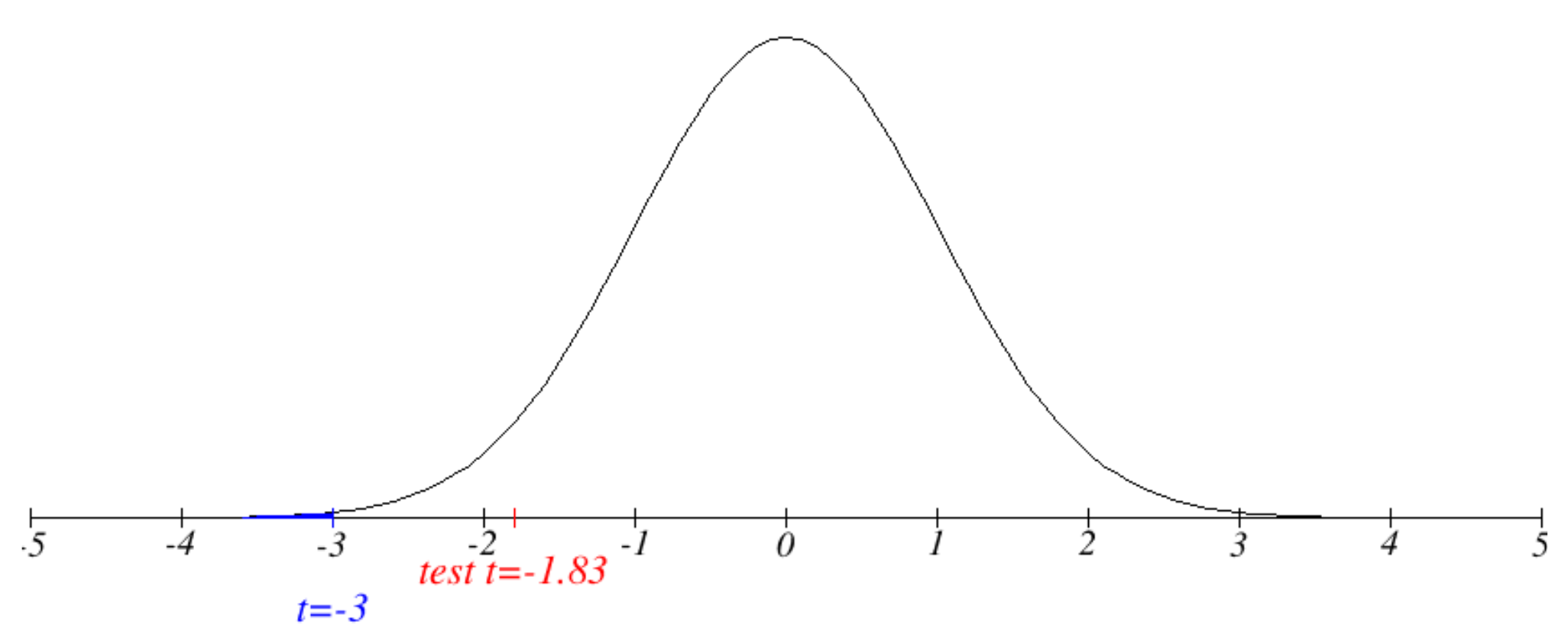}
	\caption{Hypothesis  H - RQ1 Test Graph.}
	\label{fig:test02}
\end{figure}

According to t-statistic theory, we can safely reject our null hypothesis if the \begin{math}  t-statistic\end{math} value is below the negative \begin{math} t - critical value \end{math} (threshold) \cite{peck2011statistics}. This negative \begin{math} t - critical value \end{math} bounds the area of rejection of a T-distribution, as shown in Figure \ref{fig:test02}. In our experiment, as we specified a statistical significance level of 0.01, the probability of getting a T-value less or equal than the negative  \begin{math} t - critical value \end{math} is 1\%. 
We calculated the \begin{math} t critical value \end{math}  of this T-distribution according to the T-table presented in Peck and Devore (2011, pg 791) \cite{peck2011statistics}. Accordingly, for a distribution with 7 degrees of freedom (see Table \ref{table:average}) and a confidence level of 99\%,  the negative \begin{math} t critical value \end{math}  is -3.00.
As we depicted in Figure \ref{fig:test02}, the test statistic of our sample is higher than the \begin{math} t critical value \end{math}.

As the test statistic does not fall in the critical region, we cannot safely reject this null hypothesis. Based on a t-value of -1.83 and a degree of freedom of 7, we could reject our null hypothesis  only if we had reduced the precision of our experiment to 85\%. Thus, {\bf we would fail to reject the null hypothesis and would not accept the alternative hypothesis. Therefore, we cannot state that an ML-based approach improves the performance of autonomous things compared to solutions provided by IoT expert software engineers}.

%
%

\subsubsection{How does the evaluation result from a machine learning-based solution differ from solutions provided by software engineers without IoT skills with respect to their performance?}
{\bf H - RQ3}.
\begin{itemize}
	\item H0. An ML-based approach does not improve the performance of autonomous things compared to solutions provided by software engineers without experience in IoT development.
	\item HA. An ML-based approach improves the performance of autonomous things compared to solutions provided by software engineers without experience in IoT development.
\end{itemize}

For instance, we restricted the population sample to the number of software engineers that confirmed not having experience with developing applications for the Internet of Things.  As shown in Table \ref{table:average}, the performance mean (\begin{math} \overline{x} \end{math}) of the solutions from software engineers without experience  in IoT development is 32.61 and the standard deviation (\begin{math} \sigma \end{math}) is 15.15. Thus, the 

\begin{equation}
t-statistic: t =\frac{(32.61 - 59.53)}{(\frac{15.15}{\sqrt {6}} )} = -4.35
\label{eq:test03}
\end{equation}

\begin{figure}[!htb]
	\centering
	\includegraphics[width=7.2cm]{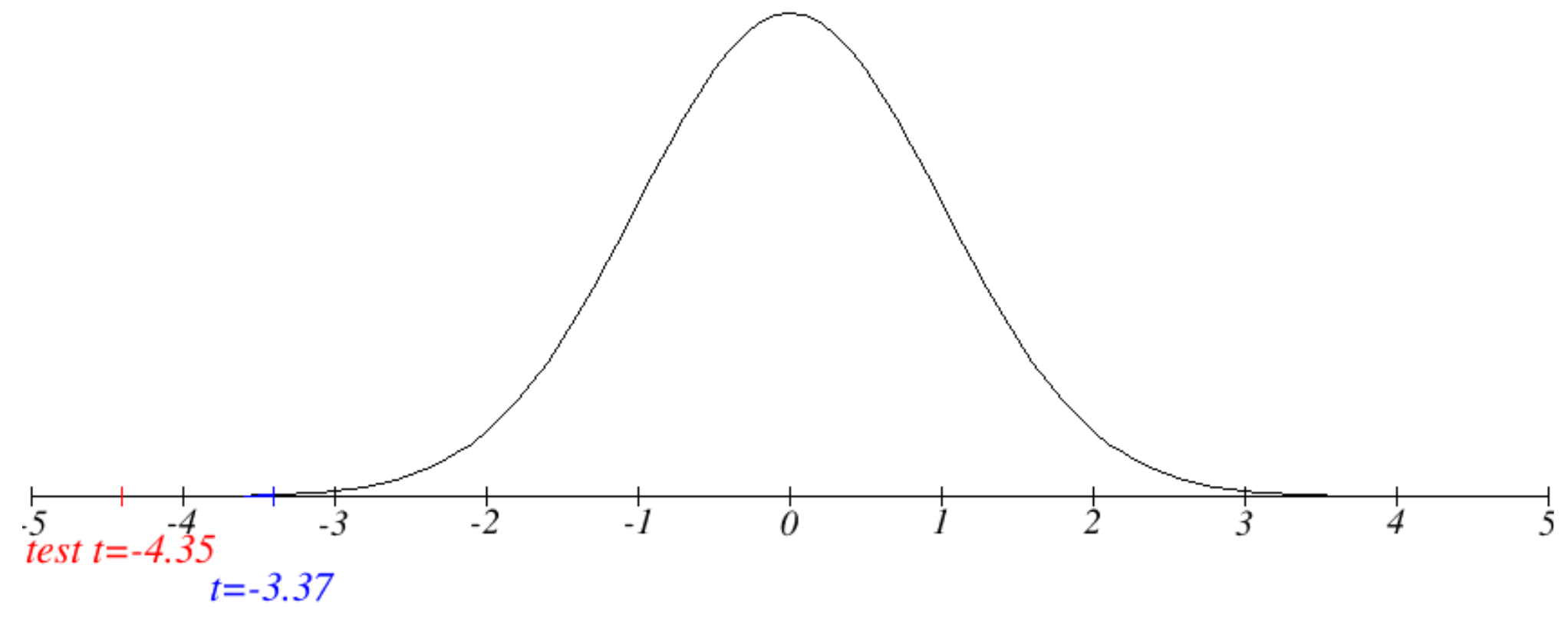}
	\caption{Hypothesis  H - RQ3 Test Graph.}
	\label{fig:test03}
\end{figure}

As shown in Table \ref{table:average}, this T-distribution has 5 degrees of freedom. Thus, for a confidence level of 99\%,  the negative \begin{math} t critical value \end{math}  is -3.37.
As we depicted in Figure \ref{fig:test03}, the test statistic of our sample is below the  \begin{math} t critical value \end{math} .

{\bf As the  \begin{math}  t-statistic\end{math} value is below the negative \begin{math} t critical value \end{math} (-4.35 $< -3.37$),  we can safely reject the null hypothesis, assuring that the error chancing of making an error is lower than 1\%}. Therefore, we accepted the alternative hypothesis: An ML-based approach improves the performance of autonomous things compared to solutions provided by software engineers with no experience in IoT development.


\section{Experiment - Part 2 - Results}

As explained previously, the second part of the experiment consists of translating the solution provided by machine learning and participants to an unknown environment. In this second part of the experiment, we also executed the simulation 16 times: for each one of the participants' solutions, for the machine-learning solution and for the zeroed solution.

Table \ref{tab:scenario2results} shows the results that were achieved by the different solutions at night in a simulation of 30 seconds. As shown, most of the solutions did not work. The person in these simulations did not finish the route even when we increased the simulation time. Only the solution provided by the machine-learning algorithm and by participant 12 worked. Remember, this scenario was not used by the machine-learning algorithm during the training process. This solution was provided through  machine learning for the first scenario and it was just reused in this new scenario. In other words, we did not restart the machine-learning process.

We selected only those solutions that worked and verified their results for the other periods of the day (morning and late afternoon). As shown in Table \ref{table:scenario2average}, when considering the whole day, the machine-learning approach presented the best result. Because the average time for the trip was a little higher using the machine-learning approach, the difference in energy consumption between the two solutions is considerably higher.

\begin{table}[!htb]
	\centering
		\caption{Using the same solution in a different environment - only at night.}
		\label{tab:scenario2results}
	\begin{tabular}{|l|l|l|l|l|}
		\hline
		\rowcolor[HTML]{000000} 
		{\color[HTML]{FFFFFF} \begin{tabular}[c]{@{}c@{}}Software\\ Engineer\end{tabular}} & {\color[HTML]{FFFFFF} Energy\%} & {\color[HTML]{FFFFFF} People\%} & {\color[HTML]{FFFFFF} Trip\%} & {\color[HTML]{FFFFFF} Fitness} \\ \hline
		1                                  & 6.50                          & 0                             & 100                         & -42.60                         \\ \hline
		2                                  & 2.77                          & 0                             & 100                         & -41.11                         \\ \hline
		3                                  & 6.62                          & 0                             & 100                         & -42.65                         \\ \hline
		4                                  & 4.30                          & 0                             & 100                         & -41.72                         \\ \hline
		5                                  & 2.58                          & 0                             & 100                         & -41.03                         \\ \hline
		6                                  & 6.88                          & 0                             & 100                         & -42.75                         \\ \hline
		7                                  & 8.33                          & 0                             & 100                         & -43.33                         \\ \hline
		8                                  & 2.33                          & 0                             & 100                         & -60.26                         \\ \hline
		9                                  & 3.77                          & 0                             & 100                         & -41.51                         \\ \hline
		10                                 & 3.78                          & 0                             & 100                         & {\color[HTML]{343434} -60.18}  \\ \hline
		11                                 & 11.36                         & 0                             & 100                         & -44.54                         \\ \hline
		\rowcolor[HTML]{656565} 
		\textbf{12}                        & \textbf{50.56}                & \textbf{100}                  & \textbf{42.22}              & {\ul \textbf{54.43} }                \\ \hline
		13                                 & 2.77                          & 0                             & 100                         & -41.11                         \\ \hline
		14                                 & 4.50                          & 0                             & 100                         & -41.80                         \\ \hline
		\rowcolor[HTML]{EFEFEF} 
		\textbf{Learning}                  & \textbf{24.44}                & \textbf{100}                  & \textbf{61.11}              & \textbf{53.55}                 \\ \hline
		zeroed                             & 0                             & 0                             & 100                         & -40                            \\ \hline
	\end{tabular}
\end{table}


\begin{table}[!htb]
	\centering
	\caption{Using the same solution in a different environment - day average.}
	\label{table:scenario2average}
	\begin{tabular}{|l|l|l|l|l|}
		\hline
		& Energy\% & People\% & Trip\%  & \textbf{Fitness}     \\ \hline
		\rowcolor[HTML]{C0C0C0} 
		\begin{tabular}[c]{@{}l@{}}Average\\ Participant\\ 12\end{tabular} & 50.52  & 100    & 38.14 & \textbf{56.90}       \\ \hline
		\begin{tabular}[c]{@{}l@{}}Average\\ Learning\end{tabular}         & 8.46   & 100    & 46.29 & {\ul \textbf{68.83}} \\ \hline
	\end{tabular}
\end{table}


\subsection{Hypothesis Testing}

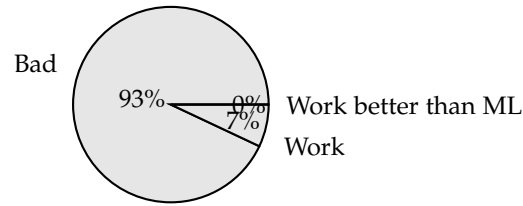
\begin{figure}[!htb]
	\begin{tikzpicture}[scale=1.3]
	
	\newcounter{g}
	\newcounter{h}
	\foreach \p/\t in {93/Bad, 7/Work, 0/Work better than ML}
	{
		\setcounter{g}{\value{h}}
		\addtocounter{h}{\p}
		\slice{\theg/100*360}
		{\theh/100*360}
		{\p\%}{\t}
	}
	
	\end{tikzpicture}
	\caption{Participants' solution results in the second scenario.}
	\label{fig:scenario2result}
\end{figure}

In this section, we investigate the hypotheses related to the solutions' reuse evaluation, that is H-RQ2 and H-RQ4, as presented in subsection \ref{sub:hyp}. Their alternative hypotheses state that an ML-based approach improves the performance of autonomous things compared to solutions provided by software engineers, software engineers with experience in IoT development, and software engineers without experience in IoT development, respectively. We planned to perform a statistical development to evaluate these hypotheses.  However, as depicted in Figure \ref{fig:scenario2result}, in the new scenario, 0\% of participants provided a result better than the result provided by the machine-learning solution. In addition, from the group of 14 engineers, only one participant, who has experience with IoT development, provided a solution that worked.


Therefore, we can safely reject the null hypothesis and accept both alternative hypotheses:

\begin{enumerate}
	\item H-RQ2: H1: An ML-based approach increases the reuse of autonomous things compared to solutions provided by IoT expert software engineers.
	\item H-RQ4: H1: 2)	An ML-based approach increases the reuse of autonomous things compared to solutions provided by software engineers with no experience with IoT development.
\end{enumerate}

\section{Discussion}
In this section, we analyze the empirical experimental results to understand which tasks are better performed by humans and which by algorithms. This is important for selecting whether software engineers or machine learning can accomplish a specific task better.

In our empirical study, in which we have assessed performance and reuse tasks, we accepted three alternative hypotheses and rejected one:

\subsubsection*{	Accepted:}
\begin{enumerate}
\item	An ML-based approach improves the performance of autonomous things compared to solutions provided by software engineers without experience with IoT development.
\item	An ML-based approach increases the reuse of autonomous things compared to solutions provided by IoT expert software engineers.
\item	An ML-based approach increases the reuse of autonomous things compared to solutions provided by software engineers without experience with IoT development.

\end{enumerate}

\subsubsection*{	Rejected:}	
\begin{enumerate}
	\item An ML-based approach improves the performance of autonomous things compared to solutions provided by IoT expert software engineers.
\end{enumerate}

Based on these results, we have found evidence that the use of machine-learning techniques can perform some SE tasks better than software engineers, considering solutions that improve performance and increase reuse. As illustrated in the experimental results, only one of the 14 software engineers provided a solution that could be reused in a new scenario. Further none of those software engineers provided a solution that works better than the ML's solution in this new scenario. If the flexibility of the application is the most important factor, based on our results, we can safely recommend the use of machine learning.

However, if we had considered performance as the only important factor to evaluate the quality of these solutions, we have found evidence that software engineers can perform SE tasks better than machine learning, considering ``better" as a solution that improves performance. As described in our experiments, we cannot state that ML improves the performance of an application in comparison to solutions provided by domain expert software engineers. This is also an interesting result as many researchers, especially in the IoT domain, have strictly focused on automating software development.

In brief, our experiment indicates that in some cases, software engineers outperform machine-learning algorithms, whereas in other cases, they do not. The evidence shows that it is important to know which one performs better in different situations in order to determine ways for software engineers to work cooperatively and effectively with automated machine-learning procedures.

\section{Threats to Validity}
Although we have designed and conducted the experiments carefully, there are always factors that can challenge the experiment’s validity. Some threats to validity as described in  \cite{wohlin2012experimentation} could indeed limit the legitimacy of our results. In this section, we present the actions taken to mitigate their impact of these factors on the research results.

As Oizumi et al. (2017) report in \cite{oizumi2017revealing}, the number of participants in the study can be a threat to validity. In addition, Fernandes et al (2016) \cite{fernandes2016information} report the diversity of participants as another possible threat. Therefore, in our study, we needed to be aware of at least two threats to validity namely: we have selected a sample of only 14 participants, which may not be enough to achieve conclusive results; and our sample consisted of only graduate students from two Brazilian universities.  Such a group may not be representative of all software engineers, who may have substantially more professional experience and background.

To mitigate the problems of the number of participants and their diversity, we selected our participants carefully. All of them have at least two years of experience with software development. In addition, we allowed participants to solve the problem by manipulating a pseudocode version, thereby avoiding gaps in the participants' knowledge, such as experience with a particular programming language or architecture. Note that  a survey was used to select participants and they all indicated a level of experience with pseudocode. The pseudocode provided by each participant was carefully translated into  Java as this is the language supported by the Framework for the Internet of Things.


Oizumi et al. (2017) reported a third threat to validity in \cite{oizumi2017revealing}, namely,  possible misunderstandings during the study. To mitigate this problem of misunderstandings, we asked all participants to write about their understanding  of the problem both before and after applying the solution. All participants  indicated that they understood the task completely. We also asked them about their confidence in their proposed solution. Most of them evaluated their own solution with the highest grade, allowing us to increase our confidence in the experimental results. In addition, we assisted the participants during the entire study, making sure they understood the experimental task




\section{Related Work}
Comparing intelligent machines to the ability of a person to solve a particular problem is not a new approach. This kind of discussion has been promoted since the beginning of Artificial Intelligence. For example, in 1997, an important moment in the history of technology happened with Garry Kasparov's 1997 chess match against the IBM supercomputer Deep Blue  \cite{kasparov2017deep}.  

 Recently, Silver et al. (2016, 2017) \cite{silver2016mastering,silver2017mastering} published a paper in the Nature Science Journal, comparing the performance of a ML technique against the results achieved by the world champion in the game of Go.  In \cite{silver2017mastering}, Silver et al. (2017) state that their program ``achieved superhuman performance."

 Whiteson et al. \cite{whiteson2005evolving} indirectly performed this comparison, by evaluating the use of three different approaches of the neuroevolution learning algorithm to solve the same tasks: (i) coevolution, that is mostly unassisted by human knowledge; (ii) layered learning, that is highly assisted; and (iii) concurrent layered learning, that is a mixed approach. The authors state that their results ``demonstrate that the appropriate level of human assistance depends critically on the difficulty of the problem.''

Furthermore, there is also a new approach in machine learning, called Automatic Machine Learning (Auto-ML) \cite{zoph2016neural}, which uses learning to set the parameters of a learning algorithm automatically. In a traditional approach, a software engineer with machine learning skills is responsible for finding a good configuration for the algorithm parameters. Zoth and Lee \cite{zoph2016neural} present an Auto-ML-based approach to design a neural network to classify images of a specific dataset. In addition, they compared their results with the previous state-of-the-art model, which was designed by an ML expert engineer. According to Zoth and Lee  \cite{zoph2016neural} , their AutoML-based approach ``can design a novel network architecture that rivals the best human-invented architecture in terms of test set accuracy." Zoth and Lee also showed that a machine-learning technique is capable of beating a software engineer with ML skills in a specific software engineering task, but the authors do not discuss this subject in the paper.



 {\bf Our paper appears to be the first to provide an empirical study to investigate the use of a machine-learning techniques to solve a problem in the field of Software Engineering, by comparing the solution provided by a ML-based approach against solutions provided by software engineers}.

\section{Conclusion and Future Work}

Several researchers have proposed the use of machine-learning techniques to automate software engineering tasks. However, most of these approaches do not direct efforts toward asking whether ML-based procedures have higher success rates than current standard and manual practices. A relevant question in this potential line of investigation is: ``Could a software engineer solve a specific development task better than an ML algorithm?". Indeed, it is fundamental to evaluate which tasks are better performed by engineers or ML procedures so that they can work together more effectively and also provide more insight into novel human-in-the-loop machine-learning approaches to support SE tasks.

This paper appears to be the first to provide an empirical study comparing how software engineers and machine-learning algorithms achieve performance and reuse tasks. In brief, as a result of our experiment, we have found evidence that in some cases, software engineers outperform machine-learning algorithms, and in other cases, they do not. Further, as is typical in experimental studies, although we have designed and conducted the experiment carefully, there are always factors that can threaten the experiment's validity.  For example, some threats include the number and diversity of the software engineers involved in our experiment.

Understanding how software engineers fare against ML algorithms is essential to support new methodologies for developing human-in-the-loop approaches in which machine learning automated procedures assist software developers in achieving their tasks. For example, methodologies to define which agent (engineers or automated ML procedure) should execute a specific task in a software development set. Based on this understanding, these methodologies can provide a basis for software engineers and machine learning algorithms to cooperate in Software Engineering development more effectively.

Future work to extend the proposed experiment includes: (i) conducting further empirical studies to assess other SE tasks, such as design, maintenance and testing; (ii) experimenting with other machine-learning algorithms such as reinforcement learning and backpropagation; and (iii) using different criteria to evaluate task execution.

Possible tasks that could be investigated (refer to (i)) include programming tasks, in which case tasks performed by software development teams and ML algorithms are compared. For example, we could invite software developers from the team with the highest score in the last ACM International Collegiate Programming Contest \cite{trotman2008programming}, which is one of the most important programming championships in the world, to be involved in this comparison. This competition evaluates the capability of software engineers to solve complex software problems. Software engineers are classified according to the number of right solutions, performance of the solutions and development time.

Another line of investigation could address the use of different qualitative or quantitative methodologies. For example, the task execution comparison could rely on reference performances, such as the performance of highly successful performers  \cite{zoph2016neural,silver2016mastering,silver2017mastering} .  This research work can also be extended by proposing, based on the comparison between the performance of engineers and ML algorithms, a methodology for more effective task allocation. This methodology could, in principle, lead to more effective ways to allocate tasks such as software development in cooperative work involving humans and automated procedures. Such human-in-the-loop approaches, which take into account the strengths and weaknesses of humans and machine learning algorithms, are fundamental to provide a basis for cooperative work in software engineering and possibly in other areas.




%


%

\ifCLASSOPTIONcompsoc
  \section*{Acknowledgments}
\else
  \section*{Acknowledgment}
\fi

		This work has been supported by the Laboratory of Software Engineering (LES) at PUC-Rio. Our thanks to CAPES, CNPq, FAPERJ and PUC-Rio for their support through scholarships and fellowships. We would also like to thank the software engineers who participated in our experiment.

\ifCLASSOPTIONcaptionsoff
  \newpage
\fi



\bibliographystyle{IEEEtran}
\bibliography{sigproc}

\begin{thebibliography}{100}
\providecommand{\url}[1]{#1}
\csname url@samestyle\endcsname
\providecommand{\newblock}{\relax}
\providecommand{\bibinfo}[2]{#2}
\providecommand{\BIBentrySTDinterwordspacing}{\spaceskip=0pt\relax}
\providecommand{\BIBentryALTinterwordstretchfactor}{4}
\providecommand{\BIBentryALTinterwordspacing}{\spaceskip=\fontdimen2\font plus
\BIBentryALTinterwordstretchfactor\fontdimen3\font minus
  \fontdimen4\font\relax}
\providecommand{\BIBforeignlanguage}[2]{{%
\expandafter\ifx\csname l@#1\endcsname\relax
\typeout{** WARNING: IEEEtran.bst: No hyphenation pattern has been}%
\typeout{** loaded for the language `#1'. Using the pattern for}%
\typeout{** the default language instead.}%
\else
\language=\csname l@#1\endcsname
\fi
#2}}
\providecommand{\BIBdecl}{\relax}
\BIBdecl

\bibitem{brooks1987no}
F.~Brooks and H.~Kugler, \emph{No silver bullet}.\hskip 1em plus 0.5em minus
  0.4em\relax April, 1987.

\bibitem{pressman2005software}
R.~S. Pressman, \emph{Software engineering: a practitioner's approach}.\hskip
  1em plus 0.5em minus 0.4em\relax Palgrave Macmillan, 2005.

\bibitem{zhang2000software}
Q.~Zhang, ``Software developments,'' \emph{Engineering Automation for Reliable
  Software}, p. 292, 2000.

\bibitem{kephart2005research}
J.~O. Kephart, ``Research challenges of autonomic computing,'' in
  \emph{Software Engineering, 2005. ICSE 2005. Proceedings. 27th International
  Conference on}.\hskip 1em plus 0.5em minus 0.4em\relax IEEE, 2005, pp.
  15--22.

\bibitem{mostow1985foreword}
J.~Mostow, ``Foreword what is ai? and what does it have to do with software
  engineering?'' \emph{IEEE Transactions on Software Engineering}, vol.~11,
  no.~11, p. 1253, 1985.

\bibitem{barstow1987artificial}
D.~Barstow, ``Artificial intelligence and software engineering,'' in
  \emph{Proceedings of the 9th international conference on Software
  Engineering}.\hskip 1em plus 0.5em minus 0.4em\relax IEEE Computer Society
  Press, 1987, pp. 200--211.

\bibitem{partridge1988artificial}
D.~Partridge, ``Artificial intelligence and software engineering: a survey of
  possibilities,'' \emph{Information and Software Technology}, vol.~30, no.~3,
  pp. 146--152, 1988.

\bibitem{cheung1991survey}
L.~C. Cheung, S.~Ip, and T.~Holden, ``Survey of artificial intelligence impacts
  on information systems engineering,'' \emph{Information and Software
  Technology}, vol.~33, no.~7, pp. 499--508, 1991.

\bibitem{partridge1998artificial}
D.~Partridge, \emph{Artificial Intelligence in Software Engineering}.\hskip 1em
  plus 0.5em minus 0.4em\relax Wiley Online Library, 1998.

\bibitem{van1998inferring}
A.~Van~Lamsweerde and L.~Willemet, ``Inferring declarative requirements
  specifications from operational scenarios,'' \emph{IEEE Transactions on
  Software Engineering}, vol.~24, no.~12, pp. 1089--1114, 1998.

\bibitem{boetticher2001using}
G.~D. Boetticher, ``Using machine learning to predict project effort: Empirical
  case studies in data-starved domains,'' in \emph{Model Based Requirements
  Workshop}.\hskip 1em plus 0.5em minus 0.4em\relax Citeseer, 2001, pp. 17--24.

\bibitem{padberg2004using}
F.~Padberg, T.~Ragg, and R.~Schoknecht, ``Using machine learning for estimating
  the defect content after an inspection,'' \emph{IEEE Transactions on Software
  Engineering}, vol.~30, no.~1, pp. 17--28, 2004.

\bibitem{zhang2000applying}
D.~Zhang, ``Applying machine learning algorithms in software development,'' in
  \emph{The Proceedings of 2000 Monterey Workshop on Modeling Software System
  Structures}, 2000, pp. 275--285.

\bibitem{zhang2006machine}
------, ``Machine learning in value-based software test data generation,'' in
  \emph{Tools with Artificial Intelligence, 2006. ICTAI'06. 18th IEEE
  International Conference on}.\hskip 1em plus 0.5em minus 0.4em\relax IEEE,
  2006, pp. 732--736.

\bibitem{zhang2005machine}
D.~Zhang and J.~J. Tsai, \emph{Machine learning applications in software
  engineering}.\hskip 1em plus 0.5em minus 0.4em\relax World Scientific, 2005,
  vol.~16.

\bibitem{zhang2008machine}
D.~Zhang, ``Machine learning and value-based software engineering: a research
  agenda.'' in \emph{SEKE}, 2008, pp. 285--290.

\bibitem{khoshgoftaar2003introduction}
T.~M. Khoshgoftaar, ``Introduction to the special issue on quality engineering
  with computational intelligence,'' 2003.

\bibitem{zhang2009machine}
D.~Zhang, ``Machine learning and value-based software engineering,'' in
  \emph{Software Applications: Concepts, Methodologies, Tools, and
  Applications}.\hskip 1em plus 0.5em minus 0.4em\relax IGI Global, 2009, pp.
  3325--3339.

\bibitem{zhang2002machine}
D.~Zhang and J.~J. Tsai, ``Machine learning and software engineering,'' in
  \emph{Tools with Artificial Intelligence, 2002.(ICTAI 2002). Proceedings.
  14th IEEE International Conference on}.\hskip 1em plus 0.5em minus
  0.4em\relax IEEE, 2002, pp. 22--29.

\bibitem{kramer2000gaps}
M.~D. Kramer and D.~Zhang, ``Gaps: a genetic programming system,'' in
  \emph{Computer Software and Applications Conference, 2000. COMPSAC 2000. The
  24th Annual International}.\hskip 1em plus 0.5em minus 0.4em\relax IEEE,
  2000, pp. 614--619.

\bibitem{holzinger2016towards}
A.~Holzinger, M.~Plass, K.~Holzinger, G.~C. Cri{\c{s}}an, C.-M. Pintea, and
  V.~Palade, ``Towards interactive machine learning (iml): applying ant colony
  algorithms to solve the traveling salesman problem with the human-in-the-loop
  approach,'' in \emph{International Conference on Availability, Reliability,
  and Security}.\hskip 1em plus 0.5em minus 0.4em\relax Springer, 2016, pp.
  81--95.

\bibitem{holzinger2016interactive}
A.~Holzinger, ``Interactive machine learning for health informatics: when do we
  need the human-in-the-loop?'' \emph{Brain Informatics}, vol.~3, no.~2, pp.
  119--131, 2016.

\bibitem{easterbrook2008selecting}
S.~Easterbrook, J.~Singer, M.-A. Storey, and D.~Damian, ``Selecting empirical
  methods for software engineering research,'' \emph{Guide to advanced
  empirical software engineering}, pp. 285--311, 2008.

\bibitem{simon1986whether}
H.~A. Simon, ``Whether software engineering needs to be artificially
  intelligent,'' \emph{IEEE Transactions on Software Engineering}, no.~7, pp.
  726--732, 1986.

\bibitem{sommerville1993artificial}
I.~Sommerville, ``Artificial intelligence and systems engineering,''
  \emph{Prospects for Artificial Intelligence: Proceedings of AISB'93, 29
  March-2 April 1993, Birmingham, UK}, vol.~17, p.~48, 1993.

\bibitem{michalski2013machine}
R.~S. Michalski, J.~G. Carbonell, and T.~M. Mitchell, \emph{Machine learning:
  An artificial intelligence approach}.\hskip 1em plus 0.5em minus 0.4em\relax
  Springer Science \& Business Media, 2013.

\bibitem{marchetto2005evaluating}
A.~Marchetto and A.~Trentini, ``Evaluating web applications testability by
  combining metrics and analogies,'' in \emph{Information and Communications
  Technology, 2005. Enabling Technologies for the New Knowledge Society: ITI
  3rd International Conference on}.\hskip 1em plus 0.5em minus 0.4em\relax
  IEEE, 2005, pp. 751--779.

\bibitem{bouktif2010novel}
S.~Bouktif, F.~Ahmed, I.~Khalil, and G.~Antoniol, ``A novel composite model
  approach to improve ality prediction,'' \emph{Information and Software
  Technology}, vol.~52, no.~12, pp. 1298--1311, 2010.

\bibitem{radlinski2010survey}
L.~Radlinski, ``A survey of bayesian net models for software development effort
  prediction,'' \emph{International Journal of Software Engineering and
  Computing}, vol.~2, no.~2, pp. 95--109, 2010.

\bibitem{zhang2011handling}
W.~Zhang, Y.~Yang, and Q.~Wang, ``Handling missing data in software effort
  prediction with naive bayes and em algorithm,'' in \emph{Proceedings of the
  7th International Conference on Predictive Models in Software
  Engineering}.\hskip 1em plus 0.5em minus 0.4em\relax ACM, 2011, p.~4.

\bibitem{radlinski2011framework}
{\L}.~Radli{\'n}ski, ``A framework for integrated software quality prediction
  using bayesian nets,'' \emph{Computational Science and Its Applications-ICCSA
  2011}, pp. 310--325, 2011.

\bibitem{sack2006building}
P.~O.~O. Sack, M.~Bouneffa, Y.~Maweed, and H.~Basson, ``On building an
  integrated and generic platform for software quality evaluation,'' in
  \emph{Information and Communication Technologies, 2006. ICTTA'06. 2nd},
  vol.~2.\hskip 1em plus 0.5em minus 0.4em\relax IEEE, 2006, pp. 2872--2877.

\bibitem{reformat2007introduction}
M.~Reformat and D.~Zhang, ``Introduction to the special issue on:``software
  quality improvements and estimations with intelligence-based methods",''
  \emph{Software Quality Journal}, vol.~15, no.~3, pp. 237--240, 2007.

\bibitem{twala2007applying}
B.~Twala, M.~Cartwright, and M.~Shepperd, ``Applying rule induction in software
  prediction,'' in \emph{Advances in Machine Learning Applications in Software
  Engineering}.\hskip 1em plus 0.5em minus 0.4em\relax IGI Global, 2007, pp.
  265--286.

\bibitem{challagulla2009high}
V.~U. Challagulla, F.~B. Bastani, and I.-L. Yen, ``High-confidence
  compositional reliability assessment of soa-based systems using machine
  learning techniques,'' in \emph{Machine Learning in Cyber Trust}.\hskip 1em
  plus 0.5em minus 0.4em\relax Springer, 2009, pp. 279--322.

\bibitem{veras2007comparative}
R.~C. Veras, S.~R. Meira, A.~L. Oliveira, and B.~J. Melo, ``Comparative study
  of clustering techniques for the organization of software repositories,'' in
  \emph{Hybrid Intelligent Systems, 2007. HIS 2007. 7th International
  Conference on}.\hskip 1em plus 0.5em minus 0.4em\relax IEEE, 2007, pp.
  372--377.

\bibitem{birzniece2010interactive}
I.~Birzniece and M.~Kirikova, ``Interactive inductive learning service for
  indirect analysis of study subject compatibility,'' in \emph{Proceedings of
  the BeneLearn}, 2010, pp. 1--6.

\bibitem{hanchate2010analysis}
D.~B. Hanchate, ``Analysis, mathematical modeling and algorithm for software
  project scheduling using bcga,'' in \emph{Intelligent Computing and
  Intelligent Systems (ICIS), 2010 IEEE International Conference on},
  vol.~3.\hskip 1em plus 0.5em minus 0.4em\relax IEEE, 2010, pp. 1--7.

\bibitem{xu2006machine}
Z.~Xu and B.~Song, ``A machine learning application for human resource data
  mining problem,'' \emph{Advances in Knowledge Discovery and Data Mining}, pp.
  847--856, 2006.

\bibitem{wen2012systematic}
J.~Wen, S.~Li, Z.~Lin, Y.~Hu, and C.~Huang, ``Systematic literature review of
  machine learning based software development effort estimation models,''
  \emph{Information and Software Technology}, vol.~54, no.~1, pp. 41--59, 2012.

\bibitem{rashid2012survey}
E.~Rashid, S.~Patnayak, and V.~Bhattacherjee, ``A survey in the area of machine
  learning and its application for software quality prediction,'' \emph{ACM
  SIGSOFT Software Engineering Notes}, vol.~37, no.~5, pp. 1--7, 2012.

\bibitem{al2013machine}
H.~A. Al-Jamimi and M.~Ahmed, ``Machine learning-based software quality
  prediction models: state of the art,'' in \emph{Information Science and
  Applications (ICISA), 2013 International Conference on}.\hskip 1em plus 0.5em
  minus 0.4em\relax IEEE, 2013, pp. 1--4.

\bibitem{radlinski2012enhancing}
{\L}.~Radli{\'n}ski, ``Enhancing bayesian network model for integrated software
  quality prediction,'' in \emph{Proc. Fourth International Conference on
  Information, Process, and Knowledge Management, Valencia}.\hskip 1em plus
  0.5em minus 0.4em\relax Citeseer, 2012, pp. 144--149.

\bibitem{pinel2013savant}
F.~Pinel, P.~Bouvry, B.~Dorronsoro, and S.~U. Khan, ``Savant: Automatic
  parallelization of a scheduling heuristic with machine learning,'' in
  \emph{Nature and Biologically Inspired Computing (NaBIC), 2013 World Congress
  on}.\hskip 1em plus 0.5em minus 0.4em\relax IEEE, 2013, pp. 52--57.

\bibitem{novitasari2016optimizing}
D.~Novitasari, I.~Cholissodin, and W.~F. Mahmudy, ``Optimizing svr using local
  best pso for software effort estimation,'' \emph{Journal of Information
  Technology and Computer Science}, vol.~1, no.~1, 2016.

\bibitem{radlinski2012towards}
{\L}.~Radli{\'n}ski, ``Towards expert-based modelling of integrated software
  quality,'' \emph{Journal of Theoretical and Applied Computer Science},
  vol.~6, no.~2, pp. 13--26, 2012.

\bibitem{rongfa2012defect}
T.~Rongfa, ``Defect classification method for software management quality
  control based on decision tree learning,'' in \emph{Advanced Technology in
  Teaching-Proceedings of the 2009 3rd International Conference on Teaching and
  Computational Science (WTCS 2009)}.\hskip 1em plus 0.5em minus 0.4em\relax
  Springer, 2012, pp. 721--728.

\bibitem{rana2015machine}
R.~Rana and M.~Staron, ``Machine learning approach for quality assessment and
  prediction in large software organizations,'' in \emph{Software Engineering
  and Service Science (ICSESS), 2015 6th IEEE International Conference
  on}.\hskip 1em plus 0.5em minus 0.4em\relax IEEE, 2015, pp. 1098--1101.

\bibitem{wang2015use}
H.~Wang, M.~Kessentini, W.~Grosky, and H.~Meddeb, ``On the use of time series
  and search based software engineering for refactoring recommendation,'' in
  \emph{Proceedings of the 7th International Conference on Management of
  computational and collective intElligence in Digital EcoSystems}.\hskip 1em
  plus 0.5em minus 0.4em\relax ACM, 2015, pp. 35--42.

\bibitem{challagulla2005empirical}
V.~U. Challagulla, F.~B. Bastani, I.-L. Yen, and R.~A. Paul, ``Empirical
  assessment of machine learning based software defect prediction techniques,''
  in \emph{Object-Oriented Real-Time Dependable Systems, 2005. WORDS 2005. 10th
  IEEE International Workshop on}.\hskip 1em plus 0.5em minus 0.4em\relax IEEE,
  2005, pp. 263--270.

\bibitem{kaminsky2004building}
K.~Kaminsky and G.~Boetticher, ``Building a genetically engineerable evolvable
  program (geep) using breadth-based explicit knowledge for predicting software
  defects,'' in \emph{Fuzzy Information, 2004. Processing NAFIPS'04. IEEE
  Annual Meeting of the}, vol.~1.\hskip 1em plus 0.5em minus 0.4em\relax IEEE,
  2004, pp. 10--15.

\bibitem{kaminsky2004predict}
------, ``How to predict more with less, defect prediction using machine
  learners in an implicitly data starved domain,'' in \emph{The 8th world
  multiconference on systemics, cybernetics and informatics, Orlando,
  FL}.\hskip 1em plus 0.5em minus 0.4em\relax Citeseer, 2004.

\bibitem{kaminsky2004better}
K.~Kaminsky and G.~D. Boetticher, ``Better software defect prediction using
  equalized learning with machine learners,'' \emph{Knowledge Sharing and
  Collaborative Engineering}, 2004.

\bibitem{kutlubay2005machine}
O.~Kutlubay and A.~Bener, ``A machine learning based model for software defect
  prediction,'' \emph{working paer, Boazi{\c{c}}i University, Computer
  Engineering Department}, 2005.

\bibitem{ren2003learn}
X.~Ren, ``Learn to predict ``affecting changes" in software engineering,''
  2003.

\bibitem{ceylan2006software}
E.~Ceylan, F.~O. Kutlubay, and A.~B. Bener, ``Software defect identification
  using machine learning techniques,'' in \emph{Software Engineering and
  Advanced Applications, 2006. SEAA'06. 32nd EUROMICRO Conference on}.\hskip
  1em plus 0.5em minus 0.4em\relax IEEE, 2006, pp. 240--247.

\bibitem{kastro2008defect}
Y.~Kastro and A.~B. Bener, ``A defect prediction method for software
  versioning,'' \emph{Software Quality Journal}, vol.~16, no.~4, pp. 543--562,
  2008.

\bibitem{kutlubay2007two}
O.~Kutlubay, B.~Turhan, and A.~B. Bener, ``A two-step model for defect density
  estimation,'' in \emph{Software Engineering and Advanced Applications, 2007.
  33rd EUROMICRO Conference on}.\hskip 1em plus 0.5em minus 0.4em\relax IEEE,
  2007, pp. 322--332.

\bibitem{namin2010bayesian}
A.~S. Namin and M.~Sridharan, ``Bayesian reasoning for software testing,'' in
  \emph{Proceedings of the FSE/SDP workshop on Future of software engineering
  research}.\hskip 1em plus 0.5em minus 0.4em\relax ACM, 2010, pp. 349--354.

\bibitem{murphy2009metamorphic}
C.~Murphy and G.~Kaiser, ``Metamorphic runtime checking of non-testable
  programs,'' \emph{Columbia University Dept of Computer Science Tech Report
  cucs-042-09}, p. 9293, 2009.

\bibitem{afzal2010genetic}
W.~Afzal, R.~Torkar, R.~Feldt, and T.~Gorschek, ``Genetic programming for
  cross-release fault count predictions in large and complex software
  projects,'' \emph{Evolutionary Computation and Optimization Algorithms in
  Software Engineering}, pp. 94--126, 2010.

\bibitem{murphy2008using}
C.~Murphy \emph{et~al.}, ``Using metamorphic testing at runtime to detect
  defects in applications without test oracles,'' 2008.

\bibitem{qiu2010framework}
D.~Qiu, S.~Fang, and Y.~Li, ``A framework to discover potential deviation
  between program and requirement through mining object graph,'' in
  \emph{Computer Application and System Modeling (ICCASM), 2010 International
  Conference on}, vol.~4.\hskip 1em plus 0.5em minus 0.4em\relax IEEE, 2010,
  pp. V4--110.

\bibitem{murphy2010automatic}
C.~Murphy, G.~E. Kaiser \emph{et~al.}, ``Automatic detection of defects in
  applications without test oracles,'' \emph{Dept. Comput. Sci., Columbia
  Univ., New York, NY, USA, Tech. Rep. CUCS-027-10}, 2010.

\bibitem{afzal2009search}
W.~Afzal, ``Search-based approaches to software fault prediction and software
  testing,'' Ph.D. dissertation, Blekinge Institute of Technology, 2009.

\bibitem{taghi2007empirical}
M.~K. Taghi, B.~Cukic, and N.~Seliya, ``An empirical assessment on program
  module-order models,'' \emph{Quality Technology \& Quantitative Management},
  vol.~4, no.~2, pp. 171--190, 2007.

\bibitem{wang2010empirical}
J.~H. Wang, N.~Bouguila, and T.~Bdiri, ``Empirical evaluation of selected
  algorithms for complexity-based classification of software modules and a new
  model,'' in \emph{Intelligent Systems: From Theory to Practice}.\hskip 1em
  plus 0.5em minus 0.4em\relax Springer, 2010, pp. 99--131.

\bibitem{jin2008artificial}
H.~Jin, Y.~Wang, N.-W. Chen, Z.-J. Gou, and S.~Wang, ``Artificial neural
  network for automatic test oracles generation,'' in \emph{Computer Science
  and Software Engineering, 2008 International Conference on}, vol.~2.\hskip
  1em plus 0.5em minus 0.4em\relax IEEE, 2008, pp. 727--730.

\bibitem{ferzund2008automated}
J.~Ferzund, S.~N. Ahsan, and F.~Wotawa, ``Automated classification of faults in
  programms using machine learning techniques,'' in \emph{Artificial
  Intelligence Techniques in Software Engineering Workshop}, 2008.

\bibitem{maqbool2007bayesian}
O.~Maqbool and H.~Babri, ``Bayesian learning for software architecture
  recovery,'' in \emph{Electrical Engineering, 2007. ICEE'07. International
  Conference on}.\hskip 1em plus 0.5em minus 0.4em\relax IEEE, 2007, pp. 1--6.

\bibitem{okutan2012software}
A.~Okutan, ``Software defect prediction using bayesian networks and kernel
  methods,'' Ph.D. dissertation, ISIK UNIVERSITY, 2012.

\bibitem{cotroneo2013learning}
D.~Cotroneo, R.~Pietrantuono, and S.~Russo, ``A learning-based method for
  combining testing techniques,'' in \emph{Proceedings of the 2013
  International Conference on Software Engineering}.\hskip 1em plus 0.5em minus
  0.4em\relax IEEE Press, 2013, pp. 142--151.

\bibitem{zhang2013value}
D.~Zhang, ``A value-based framework for software evolutionary testing,'' in
  \emph{Advances in Abstract Intelligence and Soft Computing}.\hskip 1em plus
  0.5em minus 0.4em\relax IGI Global, 2013, pp. 355--373.

\bibitem{okutan2014software}
A.~Okutan and O.~T. Y{\i}ld{\i}z, ``Software defect prediction using bayesian
  networks,'' \emph{Empirical Software Engineering}, vol.~19, no.~1, pp.
  154--181, 2014.

\bibitem{agarwal2014feature}
S.~Agarwal and D.~Tomar, ``A feature selection based model for software defect
  prediction,'' \emph{assessment}, vol.~65, 2014.

\bibitem{abaei2014important}
G.~Abaei and A.~Selamat, ``Important issues in software fault prediction: A
  road map,'' in \emph{Handbook of Research on Emerging Advancements and
  Technologies in Software Engineering}.\hskip 1em plus 0.5em minus 0.4em\relax
  IGI Global, 2014, pp. 510--539.

\bibitem{okutan2016novel}
A.~Okutan and O.~T. Yildiz, ``A novel kernel to predict software
  defectiveness,'' \emph{Journal of Systems and Software}, vol. 119, pp.
  109--121, 2016.

\bibitem{mu2012software}
X.-d. Mu, R.-h. Chang, and L.~Zhang, ``Software defect prediction based on
  competitive organization coevolutionary algorithm,'' \emph{Journal of
  Convergence Information Technology (JCIT) Volume7, Number5}, 2012.

\bibitem{cahill2013predicting}
J.~Cahill, J.~M. Hogan, and R.~Thomas, ``Predicting fault-prone software
  modules with rank sum classification,'' in \emph{Software Engineering
  Conference (ASWEC), 2013 22nd Australian}.\hskip 1em plus 0.5em minus
  0.4em\relax IEEE, 2013, pp. 211--219.

\bibitem{rana2014adoption}
R.~Rana, M.~Staron, C.~Berger, J.~Hansson, M.~Nilsson, and W.~Meding, ``The
  adoption of machine learning techniques for software defect prediction: An
  initial industrial validation,'' in \emph{Joint Conference on Knowledge-Based
  Software Engineering}.\hskip 1em plus 0.5em minus 0.4em\relax Springer, 2014,
  pp. 270--285.

\bibitem{schulz2013predicting}
T.~Schulz, {\L}.~Radli{\'n}ski, T.~Gorges, and W.~Rosenstiel, ``Predicting the
  flow of defect correction effort using a bayesian network model,''
  \emph{Empirical Software Engineering}, vol.~18, no.~3, pp. 435--477, 2013.

\bibitem{rashid2016r4}
E.~Rashid, ``R4 model for case-based reasoning and its application for software
  fault prediction,'' \emph{International Journal of Software Science and
  Computational Intelligence (IJSSCI)}, vol.~8, no.~3, pp. 19--38, 2016.

\bibitem{rashid2015improvisation}
------, ``Improvisation of case-based reasoning and its application for
  software fault prediction,'' \emph{International Journal of Services
  Technology and Management}, vol.~21, no. 4-6, pp. 214--227, 2015.

\bibitem{chhabra2014prediction}
J.~K. Chhabra and A.~Parashar, ``Prediction of changeability for object
  oriented classes and packages by mining change history,'' in \emph{Electrical
  and Computer Engineering (CCECE), 2014 IEEE 27th Canadian Conference
  on}.\hskip 1em plus 0.5em minus 0.4em\relax IEEE, 2014, pp. 1--6.

\bibitem{spanoudakis2003revising}
G.~Spanoudakis, A.~S.~d. Garcez, and A.~Zisman, ``Revising rules to capture
  requirements traceability relations: A machine learning approach.'' in
  \emph{SEKE}, 2003, pp. 570--577.

\bibitem{shin2005modeling}
M.~Shin and A.~Goel, ``Modeling software component criticality using a machine
  learning approach,'' \emph{Artificial Intelligence and Simulation}, pp.
  440--448, 2005.

\bibitem{shirabad2011predictive}
J.~S. Shirabad, ``Predictive techniques in software engineering,'' in
  \emph{Encyclopedia of Machine Learning}.\hskip 1em plus 0.5em minus
  0.4em\relax Springer, 2011, pp. 782--789.

\bibitem{araujo2016architecture}
A.~A. Ara{\'u}jo, M.~Paixao, I.~Yeltsin, A.~Dantas, and J.~Souza, ``An
  architecture based on interactive optimization and machine learning applied
  to the next release problem,'' \emph{Automated Software Engineering}, pp.
  1--49, 2016.

\bibitem{tourwe2004induced}
T.~Tourw{\'e}, J.~Brichau, A.~Kellens, and K.~Gybels, ``Induced intentional
  software views,'' \emph{Computer Languages, Systems \& Structures}, vol.~30,
  no.~1, pp. 35--47, 2004.

\bibitem{di2002machine}
J.~S. Di~Stefano and T.~Menzies, ``Machine learning for software engineering:
  Case studies in software reuse,'' in \emph{Tools with Artificial
  Intelligence, 2002.(ICTAI 2002). Proceedings. 14th IEEE International
  Conference on}.\hskip 1em plus 0.5em minus 0.4em\relax IEEE, 2002, pp.
  246--251.

\bibitem{fu2006automated}
J.~Fu, F.~B. Bastani, and I.-L. Yen, ``Automated ai planning and code pattern
  based code synthesis,'' in \emph{Tools with Artificial Intelligence, 2006.
  ICTAI'06. 18th IEEE International Conference on}.\hskip 1em plus 0.5em minus
  0.4em\relax IEEE, 2006, pp. 540--546.

\bibitem{fu2010semantic}
J.~Fu, F.~B. Bastani, I.-L. YEN \emph{et~al.}, ``Semantic-driven
  component-based automated code synthesis,'' \emph{Semantic Computing}, pp.
  249--283, 2010.

\bibitem{katasonov2008smart}
A.~Katasonov, O.~Kaykova, O.~Khriyenko, S.~Nikitin, and V.~Y. Terziyan, ``Smart
  semantic middleware for the internet of things.'' \emph{ICINCO-ICSO}, vol.~8,
  pp. 169--178, 2008.

\bibitem{baresi2014short}
L.~Baresi, S.~Guinea, and A.~Shahzada, ``Short paper: Harmonizing heterogeneous
  components in sesame,'' in \emph{Internet of Things (WF-IoT), 2014 IEEE World
  Forum on}.\hskip 1em plus 0.5em minus 0.4em\relax IEEE, 2014, pp. 197--198.

\bibitem{zhu2014minson}
L.~Zhu, H.~Cai, and L.~Jiang, ``Minson: A business process self-adaptive
  framework for smart office based on multi-agent,'' in \emph{e-Business
  Engineering (ICEBE), 2014 IEEE 11th International Conference on}.\hskip 1em
  plus 0.5em minus 0.4em\relax IEEE, 2014, pp. 31--37.

\bibitem{de2016intelligent}
J.~F. De~Paz, J.~Bajo, S.~Rodr{\'\i}guez, G.~Villarrubia, and J.~M. Corchado,
  ``Intelligent system for lighting control in smart cities,''
  \emph{Information Sciences}, vol. 372, pp. 241--255, 2016.

\bibitem{birzniece2010use}
I.~Birzniece, ``The use of inductive learning in information systems,'' in
  \emph{Proceedings of the 16th International Conference on Information and
  Software Technologies}, 2010, pp. 95--101.

\bibitem{alrajeh2006inferring}
D.~Alrajeh, A.~Russo, and S.~Uchitel, ``Inferring operational requirements from
  scenarios and goal models using inductive learning,'' in \emph{Proceedings of
  the 2006 international workshop on Scenarios and state machines: models,
  algorithms, and tools}.\hskip 1em plus 0.5em minus 0.4em\relax ACM, 2006, pp.
  29--36.

\bibitem{sharifloo2016learning}
A.~M. Sharifloo, A.~Metzger, C.~Quinton, L.~Baresi, and K.~Pohl, ``Learning and
  evolution in dynamic software product lines,'' in \emph{Proceedings of the
  11th International Symposium on Software Engineering for Adaptive and
  Self-Managing Systems}.\hskip 1em plus 0.5em minus 0.4em\relax ACM, 2016, pp.
  158--164.

\bibitem{zoph2016neural}
B.~Zoph and Q.~V. Le, ``Neural architecture search with reinforcement
  learning,'' \emph{arXiv preprint arXiv:1611.01578}, 2016.

\bibitem{do2017fiot}
N.~M. do~Nascimento and C.~J.~P. de~Lucena, ``Fiot: An agent-based framework
  for self-adaptive and self-organizing applications based on the internet of
  things,'' \emph{Information Sciences}, vol. 378, pp. 161--176, 2017.

\bibitem{jacob2010code}
F.~Jacob and R.~Tairas, ``Code template inference using language models,'' in
  \emph{Proceedings of the 48th Annual Southeast Regional Conference}.\hskip
  1em plus 0.5em minus 0.4em\relax ACM, 2010, p. 104.

\bibitem{amal2014use}
B.~Amal, M.~Kessentini, S.~Bechikh, J.~Dea, and L.~B. Said, ``On the use of
  machine learning and search-based software engineering for ill-defined
  fitness function: a case study on software refactoring,'' in
  \emph{International Symposium on Search Based Software Engineering}.\hskip
  1em plus 0.5em minus 0.4em\relax Springer, 2014, pp. 31--45.

\bibitem{peng2012user}
Y.~Peng, G.~Wang, and H.~Wang, ``User preferences based software defect
  detection algorithms selection using mcdm,'' \emph{Information Sciences},
  vol. 191, pp. 3--13, 2012.

\bibitem{abbass2016trusted}
H.~A. Abbass, E.~Petraki, K.~Merrick, J.~Harvey, and M.~Barlow, ``Trusted
  autonomy and cognitive cyber symbiosis: Open challenges,'' \emph{Cognitive
  computation}, vol.~8, no.~3, pp. 385--408, 2016.

\bibitem{baxt1991use}
W.~G. Baxt, ``Use of an artificial neural network for the diagnosis of
  myocardial infarction,'' \emph{Annals of internal medicine}, vol. 115,
  no.~11, pp. 843--848, 1991.

\bibitem{mazurowski2008training}
M.~A. Mazurowski, P.~A. Habas, J.~M. Zurada, J.~Y. Lo, J.~A. Baker, and G.~D.
  Tourassi, ``Training neural network classifiers for medical decision making:
  The effects of imbalanced datasets on classification performance,''
  \emph{Neural networks}, vol.~21, no.~2, pp. 427--436, 2008.

\bibitem{doiot}
N.~M. do~Nascimento, M.~L. Viana, and C.~J.~P. de~Lucena, ``An iot-based tool
  for human gas monitoring,'' in \emph{IXV Congresso Brasileiro de Informatica
  em Saude (CBIS)}, vol.~1.\hskip 1em plus 0.5em minus 0.4em\relax SBIS, 2016,
  pp. 96--98.

\bibitem{morejon2017generating}
R.~Morej{\'o}n, M.~Viana, and C.~Lucena, ``Generating software agents for data
  mining: An example for the health data area,'' in \emph{International
  Conference on Software Engineering \& Knowledge Engineering-SEKE}, 2017.

\bibitem{haykin1994neural}
\BIBentryALTinterwordspacing
S.~Haykin, \emph{Neural Networks: A Comprehensive Foundation}.\hskip 1em plus
  0.5em minus 0.4em\relax Macmillan, 1994. [Online]. Available:
  \url{http://books.google.com.br/books?id=PSAPAQAAMAAJ}
\BIBentrySTDinterwordspacing

\bibitem{castelvecchi2016can}
D.~Castelvecchi, ``Can we open the black box of ai?'' \emph{Nature News}, vol.
  538, no. 7623, p.~20, 2016.

\bibitem{atzori2010internet}
L.~Atzori, A.~Iera, and G.~Morabito, ``The internet of things: A survey,''
  \emph{Computer networks}, vol.~54, no.~15, pp. 2787--2805, 2010.

\bibitem{salahuddin2017softwarization}
M.~A. Salahuddin, A.~Al-Fuqaha, M.~Guizani, K.~Shuaib, and F.~Sallabi,
  ``Softwarization of internet of things infrastructure for secure and smart
  healthcare,'' \emph{Computer}, vol.~50, no.~7, pp. 74--79, 2017.

\bibitem{ayala2015software}
I.~Ayala, M.~Amor, L.~Fuentes, and J.~M. Troya, ``A software product line
  process to develop agents for the iot,'' \emph{Sensors}, vol.~15, no.~7, pp.
  15\,640--15\,660, 2015.

\bibitem{briot2016multi}
J.-P. Briot, N.~M. de~Nascimento, and C.~J.~P. de~Lucena, ``A multi-agent
  architecture for quantified fruits: Design and experience,'' in \emph{28th
  International Conference on Software Engineering \& Knowledge Engineering
  (SEKE'2016)}.\hskip 1em plus 0.5em minus 0.4em\relax SEKE/Knowledge Systems
  Institute, PA, USA, 2016, pp. 369--374.

\bibitem{nathalia:mestrado:15}
N.~M. Nascimento, ``{FIoT}: An agent-based framework for self-adaptive and
  self-organizing internet of things applications,'' Master's thesis, PUC-Rio,
  Rio de Janeiro, Brazil, August 2015.

\bibitem{nascimento2015modeling}
N.~M.~d. Nascimento, C.~J. P.~d. Lucena, and H.~Fuks, ``Modeling quantified
  things using a multi-agent system,'' in \emph{IEEE / WIC / ACM International
  Conference on Web Intelligence and Intelligent Agent Technology (WI-IAT)},
  vol.~1.\hskip 1em plus 0.5em minus 0.4em\relax IEEE, 2015, pp. 26--32.

\bibitem{nascimento2017engineering}
N.~M. NASCIMENTO and C.~J.~P. LUCENA, ``Engineering cooperative smart things
  based on embodied cognition,'' in \emph{NASA/ESA Conference on Adaptive
  Hardware and Systems (AHS 2017)}.\hskip 1em plus 0.5em minus 0.4em\relax
  IEEE, 2017.

\bibitem{homekit}
Apple, ``Homekit,'' https://developer.apple.com/homekit/, March 2017.

\bibitem{smartthingssamsung}
Samsung, ``Samsung smart things,'' https://www.smartthings.com, March 2017.

\bibitem{sjoberg2008building}
D.~I. Sj{\o}berg, T.~Dyb{\aa}, B.~C. Anda, and J.~E. Hannay, ``Building
  theories in software engineering,'' \emph{Guide to advanced empirical
  software engineering}, pp. 312--336, 2008.

\bibitem{whiteson2005evolving}
S.~Whiteson, N.~Kohl, R.~Miikkulainen, and P.~Stone, ``Evolving soccer keepaway
  players through task decomposition,'' \emph{Machine Learning}, vol.~59, no.
  1-2, pp. 5--30, 2005.

\bibitem{wohlin2012experimentation}
C.~Wohlin, P.~Runeson, M.~H{\"o}st, M.~C. Ohlsson, B.~Regnell, and
  A.~Wessl{\'e}n, \emph{Experimentation in software engineering}.\hskip 1em
  plus 0.5em minus 0.4em\relax Springer Science \& Business Media, 2012.

\bibitem{sutton1998reinforcement}
R.~S. Sutton and A.~G. Barto, \emph{Reinforcement learning: An
  introduction}.\hskip 1em plus 0.5em minus 0.4em\relax MIT press Cambridge,
  1998, vol.~1, no.~1.

\bibitem{peck2011statistics}
R.~Peck and J.~Devore, \emph{Statistics: The Exploration \& Analysis of
  Data}.\hskip 1em plus 0.5em minus 0.4em\relax Nelson Education, 2011.

\bibitem{oizumi2017revealing}
W.~Oizumi, L.~Sousa, A.~Garcia, R.~Oliveira, A.~Oliveira, O.~Agbachi, and
  C.~Lucena, ``Revealing design problems in stinky code: a mixed-method
  study,'' in \emph{Proceedings of the 11th Brazilian Symposium on Software
  Components, Architectures, and Reuse}.\hskip 1em plus 0.5em minus 0.4em\relax
  ACM, 2017, p.~5.

\bibitem{fernandes2016information}
E.~Fernandes, F.~Ferreira, J.~A. Netto, and E.~Figueiredo, ``Information
  systems development with pair programming: An academic quasi-experiment,'' in
  \emph{Proceedings of the XII Brazilian Symposium on Information Systems on
  Brazilian Symposium on Information Systems: Information Systems in the Cloud
  Computing Era-Volume 1}.\hskip 1em plus 0.5em minus 0.4em\relax Brazilian
  Computer Society, 2016, p.~64.

\bibitem{kasparov2017deep}
G.~Kasparov, \emph{Deep Thinking: Where Machine Intelligence Ends and Human
  Creativity Begins}.\hskip 1em plus 0.5em minus 0.4em\relax Hachette UK, 2017.

\bibitem{silver2016mastering}
D.~Silver, A.~Huang, C.~J. Maddison, A.~Guez, L.~Sifre, G.~Van Den~Driessche,
  J.~Schrittwieser, I.~Antonoglou, V.~Panneershelvam, M.~Lanctot \emph{et~al.},
  ``Mastering the game of go with deep neural networks and tree search,''
  \emph{Nature}, vol. 529, no. 7587, pp. 484--489, 2016.

\bibitem{silver2017mastering}
D.~Silver, J.~Schrittwieser, K.~Simonyan, I.~Antonoglou, A.~Huang, A.~Guez,
  T.~Hubert, L.~Baker, M.~Lai, A.~Bolton \emph{et~al.}, ``Mastering the game of
  go without human knowledge,'' \emph{Nature}, vol. 550, no. 7676, pp.
  354--359, 2017.

\bibitem{trotman2008programming}
A.~Trotman and C.~Handley, ``Programming contest strategy,'' \emph{Computers \&
  Education}, vol.~50, no.~3, pp. 821--837, 2008.

\end{thebibliography}

\end{document}